\documentclass{article}
\usepackage{color}
\usepackage{graphicx}
\usepackage{amsmath, amsthm}
\usepackage{mathtools}
\usepackage{url}
\usepackage{soulutf8}
\RequirePackage[colorlinks,citecolor=blue,urlcolor=blue]{hyperref}
\usepackage{xcolor}
\usepackage{chngcntr}
\usepackage{subfig}
\usepackage{subcaption} 
\usepackage{booktabs}
\usepackage{tikz}
\usepackage{makecell}
\usepackage{hyperref}
\usepackage{algorithm}
\usepackage[noend]{algpseudocode}
\usepackage{setspace}

\usepackage{xfrac}
\definecolor{forest}{rgb}{0.133,0.545,0.133}
\usepackage{multirow}
\usepackage{amsfonts}

\evensidemargin=\oddsidemargin
\usepackage{etoolbox}

\newif\ifabbreviation
\pretocmd{\thebibliography}{\abbreviationfalse}{}{}
\AtBeginDocument{\abbreviationtrue}

\begin{document}
	\newcommand{\bb}{\boldsymbol{\beta}}

	\title{A General Mixture Loss Function to Optimize a Personalized Predictive Model}


	\author{Tatiana Krikella$^1$\footnote{Tatiana Krikella is the corresponding author and may be contacted at \url{krikella@yorku.ca}.} \hspace{35pt} Joel A. Dubin$^2$ \bigskip \\  
    \textit{$^1$Department of Mathematics \& Statistics} \\ \textit{York University, Toronto, ON, Canada, M3J 1P3} \\ 
 \\
    \textit{$^2$Department of Statistics \& Actuarial Science} \\ \textit{University of Waterloo, Waterloo, ON, Canada, N2L 3G1}}  
    
	\date{}

	\maketitle

	\begin{abstract}

Advances in precision medicine increasingly drive methodological innovation in health research. A key development is the use of personalized prediction models (PPMs), which are fit using a similar subpopulation tailored to a specific index patient, and have been shown to outperform one-size-fits-all models, particularly in terms of model discrimination performance. We propose a generalized loss function that enables tuning of the subpopulation size used to fit a PPM. This loss function allows joint optimization of discrimination and calibration, allowing both the performance measures and their relative weights to be specified by the user. To reduce computational burden, we conducted extensive simulation studies to identify practical bounds for the grid of subpopulation sizes. Based on these results, we recommend using a lower bound of 20\% and an upper bound of 70\% of the entire training dataset. We apply the proposed method to both simulated and real-world datasets and demonstrate that previously observed relationships between subpopulation size and model performance are robust. Furthermore, we show that the choice of performance measures in the loss function influences the optimal subpopulation size selected. These findings support the flexible and computationally efficient implementation of PPMs in precision health research.
\end{abstract}

		\bigskip


	\maketitle

	\baselineskip=19.5pt


    \section{Introduction}
\label{s:intro}

Personalized predictive models (PPMs) are models which make predictions for a patient of interest, whom we call the index patient, using a model developed on a subpopulation of patients similar to that index patient. It has been shown that PPMs can lead to better prediction for an index patient compared to one-size-fits-all models which instead use the entire available population as training data \cite{celi2012database, lee2015personalized, krikella2025personalized}. When fitting a PPM, one needs to decide how to define similarity, how many patients to include in the similar subpopulation, and how to measure model performance. Focusing on a binary response prediction problem and motivated by the fact that it has been criticized that calibration is not assessed nearly as often as discrimination \cite{van2019calibration}, in previous work \cite{krikella2025personalized} we proposed an algorithm to fit a PPM which jointly optimizes discrimination and calibration. \textit{Discrimination} is the ability of the predictive model to distinguish between outcome classes, while \textit{calibration} is the level of agreement between the observed outcomes and the predictions.

In our published algorithm, the size of subpopulation is tuned to minimize a loss function, 
which is an extension of a decomposition of the Brier Score, an overall performance measure, that consists of terms related to both discrimination and calibration. 
Specifically, the measure of calibration is related to Spiegelhalter's z-statistic, and the measure of discrimination measures the lack of spread in the predictions \cite{spiegelhalter1986probabilistic}. Both of these terms have an expected value of 0 under perfect performance, and we include a mixture term $\alpha \in [0,1]$ with the purpose of allowing the emphasis of one performance measure over another based on the needs of the researcher. 



\cite{van2016calibration} propose a calibration hierarchy consisting of four increasingly strict levels of calibration: mean, weak, moderate and strong. They say that although strong calibration is the utopic goal, it is unrealistic in practice, thus we should strive for models that are moderately calibrated. Many calibration measures reported in the literature do not reach the level of moderate calibration. Thus, we want to use a moderate measure of calibration when possible, such as the Integrated Calibration Index \cite{austin2019integrated}. 

Many algorithms proposed to fit PPMs have been based on optimizing the size of subpopulation using a popular measure of discrimination, the area under the receiver operating characteristic curve (AUROC) \cite{ng2015personalized, wang2021study, campillo2013improving, lee2015personalized}. The relationship between the size of subpopulation, $M$, and the AUROC has thus been earlier investigated and,
in \cite{krikella2025personalized}, we confirmed the results of \cite{lee2015personalized} and demonstrated that the relationship between the AUROC and $M$ is negatively proportional, aside from the lowest values of $M$. We do not know whether this relationship is specific to the data considered: in \cite{lee2015personalized}, only one dataset was used, and in \cite{krikella2025personalized}, we only considered one simulated dataset and one dataset from the eICU Collaborative Research (intensive care unit patient) database \cite{pollard2018eicu}. Further, we showed that although the best calibrated models use the entire training data, a smaller $M$ may be the better choice that still results in a well-calibrated PPM. We want to further explore this relationship between $M$ and calibration to determine whether this is much more generally the case.

We wish to address some important limitations of earlier our algorithm \cite{krikella2025personalized}. Specifically, we propose an updated loss function to tune $M$ that is more generalized and allows the user to choose the measures of discrimination and calibration in the function. We also test our algorithm under a variety of datasets to further investigate the relationship between $M$ and a number of performance measures, the results of which we then use to define lower and upper bounds on the grid of values on which to tune $M$ to reduce the computational burden of the algorithm. 

This paper is organized as follows. We propose the updated loss function, along with an updated algorithm, in Section 2. This algorithm is then tested on 12 simulated datasets to determine the robustness of the relationship between $M$ and various performance measures, the design of which and results are included in Section 3. Based on these results, we also provide recommendations for obtaining lower and upper bounds for the grid of values on which to tune $M$. In Section 4, our proposed updated algorithm is tested on data from the eICU database\cite{pollard2018eicu} using two different loss functions, each differing in the measure of discrimination chosen. We conclude with a discussion of this proposed work in Section 5.

\section{Methods}
\label{s:methods} 

\subsection{Notation} 

Let there be $n_{train}$ patients in the training data, where each patient $k$ is represented by a Euclidean vector, $\boldsymbol{X}_k$, $k$ = 1, ..., $n_{train}$, in the multi-dimension predictor space defined by $p$ predictors. 
We let $x_h$, $h = 1,...,p$, represent the $n_{train} \times 1$ vector consisting of each patient's value of a given predictor. 
Further, let $\boldsymbol{X}_I$, $I = 1, ..., n_{test}$ be the Euclidean vector representation of an index patient, $I$, in the test data for whom we would like to make a prediction.

The focus of the prediction task in this paper is on binary classification. Thus, we consider a binary outcome. 
The actual outcome for an index patient $I$ is represented by $y_I \in \{0,1\}$.
Let $M$ be a scalar representing the size of the subpopulation in the training data consisting of similar patients to the index patient.  
Finally, let $p_I^{(M)} \in [0,1]$ represent the predicted probability of the outcome for a patient $I$ found using a model fit on a similar subpopulation of size $M$.

\subsection{Patient Similarity Metric} 

Prior to fitting a PPM, one must decide on a patient similarity metric to use. That is, a criterion that will be used to measure the similarity between individuals. In general, we label this similarity metric as $d(\boldsymbol{X}_I, \boldsymbol{X}_k)$, which represents the similarity between an index patient $I$ and a patient $k$ in the training set.  In our work, we focus on an unsupervised process, and ideally use  subject-matter specific covariates that will be used both for the similarity metric calculation and in the prediction models. In our context, the unsupervised method is appropriate, since, in some settings, such as a new patient $I$ being admitted to a hospital, we will not have an outcome label. This said, in much of our work in this paper, we will have an outcome label for those in the test data in order to evaluate predictive model performance.

We use the same patient similarity metric as was used in \cite{krikella2025personalized}: the \textit{cosine similarity metric}. This was chosen as it is bounded, interpretable and able to handle non-continuous data. For more details, please see \cite{krikella2025personalized}.

\subsection{Performance Measures}
\label{ss:performance}

\label{ss: performance2}

To measure the performance of a PPM, one should measure both the discrimination and calibration of the model \cite{steyerberg_2019}. In this study, to measure discrimination we use the area under the receiver operating characteristic curve (AUROC), the area under the precision recall curve (AUPRC), and the \textit{lack of spread} in predictions. 

The AUROC is a common measure of discrimination, however, in cases where the outcome we wish to predict is rare, ie. has a low prevalence, the AUROC is a better measure of discrimination \cite{davis2006relationship, ozenne2015precision}. The precision recall curve plots the trade-off between precision, which is $\frac{\text{true positive}}{\text{true positive + false positive}}$, often referred to as the \textit{positive predictive value} (or PPV), and sensitivity, for all possible thresholds. The AUPRC has a value between 0 and 1, but unlike the AUROC which will be 0.5 for an uninformative model, the AUPRC of an uninformative model is dependent on the prevalence, and approaches 0 as the prevalence decreases.

The lack of spread in predictions \cite{spiegelhalter1986probabilistic}, although not used often in isolation in practice, is the measure corresponding to discrimination in the Brier Score decomposition discussed by \cite{rufibach2010use}. This term is 0 when the discrimination is perfect (ie. all predictions are either 1 or 0), and under the worst case scenario (ie. all $p_I^{(M)}=0.5$), the value of this term is 0.25.

We consider measures from the three increasingly strict levels of calibration proposed by \cite{van2016calibration}. These measures, in order of strength, include: the calibration-in-the-large (CITL), the calibration slope, and the integrated calibration index (ICI) \cite{steyerberg_2019}. These measures are identical to those used as measure of calibration in our previous work. For more details on these measures, please see \cite{krikella2025personalized}.


\cite{rufibach2010use} has introduced a two-term decomposition of the Brier Score into its calibration and discrimination subparts, shown in Equation (\ref{eq: BrierDecomp}) below. Here, $n_{test}$ denotes the size of the testing set (ie. the number of index patients to which we are fitting a PPM), $y_I$ corresponds to the observed binary outcome of the index patient, $I$, and $\hat{p}_I^{(M)}$ corresponds to the predicted probability for individual $I$ experiencing the event, obtained from fitting a PPM on a similar subpopulation of size $M$.

\begin{equation}
  \label{eq: BrierDecomp}
\frac{1}{n_{test}} \sum_{I=1}^{n_{test}} (y_I - \hat{p}_I^{(M)})^2 
= \frac{1}{n_{test}} \sum_{I=1}^{n_{test}} [(y_I - \hat{p}_I^{(M)})(1-2\hat{p}_I^{(M)})] 
+ \frac{1}{n_{test}} \sum_{I=1}^{n_{test}} [\hat{p}_I^{(M)}(1-\hat{p}_I^{(M)})]  
\end{equation}

The first term after the equal sign measures the lack of calibration, with an expectation of 0 under perfect calibration. Specifically, if we define the calibration in terms of the calibration hierarchy, we notice that since this term measures the overall agreement between the predicted probabilities to the actual outcomes, this term is actually a measure of mean calibration, the weakest measure. The second term after the equal sign, as mentioned above, is a measure of discrimination, specifically, it is the lack of spread of the predictions. 

\cite{van2016calibration} recommend that when fitting clinical predictions models, we strive for moderate calibration, however the Brier Score assesses calibration through mean calibration (and discrimination through lack of spread). Further, it has been shown that the Brier Score may select a model that is miscalibrated with good discrimination over a model that is better calibrated with average discrimination \cite{assel2017brier}. These facts support that we should be assessing discrimination and calibration separately to obtain a better representation of a model's performance.

\subsection{Tuning the Size of Subpopulation}
\label{ss: tuningAlgorithm}

When fitting a PPM, \cite{lee2017patient} suggests that choosing the optimal size of subpopulation, $M$, will become important when taking these methods into practice. Motivated by this idea, we introduced an algorithm (\cite{krikella2025personalized}) that tunes $M$ to find the value that jointly optimizes discrimination and calibration using a loss function, shown in Equation (\ref{eq: introLoss}) that is inspired by \cite{rufibach2010use}'s two-term decomposition of the Brier Score: 

\begin{equation}
\label{eq: introLoss}
    L^{(M)} = \frac{\alpha}{n_{test}}\sum_{I=1}^{n_{test}}(y_I - \hat{p}_I^{(M)})(1 - 2\hat{p}_I^{(M)})
 + \frac{1-\alpha}{n_{test}}\sum_{I=1}^{n_{test}}\hat{p}_I^{(M)}(1-\hat{p}_I^{(M)}).
\end{equation}

The difference between our loss function in Equation (\ref{eq: introLoss}) and the Brier Score decomposition in Equation (\ref{eq: BrierDecomp}) is that we have included a mixture term, $\alpha$, which allows the emphasis of one performance measure over the other, if desired. As we increase $\alpha$ in Equation (\ref{eq: introLoss}), more emphasis is put on the calibration term in the loss function. Using this loss function is acceptable, especially if mean, weak and moderate calibration will always result in the same optimal $M$ value. However, because we do not know this for sure, we propose another method of tuning the size of subpopulation. We present a more general loss function that allows the user to choose their preferred measures of calibration and discrimination, represented by $C$ and $D$, respectively. This general loss function is shown in Equation (\ref{eq: generalLoss}), and note that we keep the mixture term $\alpha \in \left[0, 1\right]$, which allows flexibility in weighting one performance measure over another. Alternatively, a researcher can decide to tune alpha.
 
\begin{equation}
\label{eq: generalLoss} 
    L^{(M)}= \frac{\alpha}{n_{test}}\sum_{I=1}^{n_{test}} C_k^{(M)} + \frac{1-\alpha}{n_{test}}\sum_{I=1}^{n_{test}} D_k^{(M)}
\end{equation}

We can, for example, use the ICI as the calibration measure, and lack of spread as the discrimination measure, to obtain the loss function shown in Equation (\ref{eq: newLoss}), which optimizes $M$ in terms of discrimination and moderate calibration. In Equation (\ref{eq: newLoss}), $\tilde{p}_I^{(M)}$ are the smoothed predicted probabilities based on the LOESS calibration curve:

\begin{equation}
\label{eq: newLoss}
    L^{*(M)} = \frac{\alpha}{n_{test}}\sum_{I=1}^{n_{test}}\left|\tilde{p}_I^{(M)} - \hat{p}_I^{(M)}\right| + \frac{1-\alpha}{n_{test}}\sum_{I=1}^{n_{test}}\hat{p}_I^{(M)}(1-\hat{p}_I^{(M)}).
\end{equation}

The original iteration of this algorithm \cite{krikella2025personalized} also suffered from a large computational burden. One way of remedying this computational burden we are proposing in this paper is to determine a lower and upper bound on the grid of values used to tune $M$. Specifying a lower bound would also rectify the limitation of the algorithm where an optimal $M$ proportion could be chosen in the training/testing step that is not a sufficient sample size in the validation step.

To determine appropriate bounds for $M$, we must assess the robustness of its relationship with performance measures described in \cite{krikella2025personalized}. This ensures that the selected bounds preserve both discrimination and calibration quality. Prior findings indicate that lower values of $M$ yield good discrimination, while the relationship between $M$ and calibration is generally quadratic. If these trends are robust across datasets with varying characteristics, as evaluated in Section 3, we will choose the lower bound to ensure sufficient sample size, and the upper bound to balance improved calibration without substantial loss in discrimination. The recommended bounds, in the form of proportions, are formally introduced following the simulation results in Section 3.2.

An updated algorithm is proposed below, with the purpose of both offering an improved way of finding the optimal size of subpopulation and lessening the computational burden.

\begin{enumerate}
     \item Split the entire population into a $(100q)$\% hold-out validation set ($n_{val}$) and $(100[1-q])$\% training-testing (TrTe) set ($n_{TrTe}$), where $q\in[0,1]$. We often use $q = 0.2$.
    \item Determine a measure of discrimination and a measure of calibration to be included in the generalized loss function, $L^{(M)}$, displayed in Equation (\ref{eq: generalLoss}), along with a desired $\alpha$ value.
    \item Tune $M$ by following the steps below:
    \begin{enumerate}
        \item Randomly split the TrTe set into distinct training and testing sets using $K$-fold cross validation.
        \begin{enumerate}
            \item For index patient $I$ in the test set, calculate $d(\boldsymbol{X}_I, \boldsymbol{X}_k)$, for $k=1,...,n_{train}$ in the training set, where $d(\boldsymbol{X}_I, \boldsymbol{X}_k)$ is the similarity metric between the index patient and the $k^{th}$ training set patient.
            \item Using our recommendations for an upper and lower bound described in Section 3, define a grid of $M$ values and tune $M$ through random grid search. 
            \begin{enumerate}
                \item Sort all $d(\boldsymbol{X}_I, \boldsymbol{X}_k)$ in descending order, and the patients corresponding to the top $M$ scores form the subpopulation.
                \item Fit the predictive model of interest, such as logistic regression, random forest, etc. on the subpopulation.
                \item Average the values of the specified loss function over the $K$-folds.
            \end{enumerate} 
            \item Repeat the $K$-fold cross validation $v$ times, where $vK \geq 200$.
            \item The optimal $M$, called $M^{opt}$, is chosen as the one which minimizes the updated loss function over the $v$ repeated $K$-folds.
            \item $M^{opt}$ is divided by the number of patients in the training set to obtain the optimal $M$ proportion, $M_{prop}^{opt} = M^{opt}/n_{train}$.
        \end{enumerate}
    \end{enumerate} 
        \item Draw $B$ bootstrap samples from the validation sample (each the same size as the validation sample). Do the following steps for each bootstrap sample:
        \begin{enumerate}
            \item Randomly split the bootstrap sample into an 80\% training sample and 20\% testing sample. 
            \item Set the integer $M = ceiling(n_{val.train}M_{prop}^{opt})$ where $n_{val.train}$ is the size of the training set.  
            \item For index patient $I$ in the test set, calculate $d(\boldsymbol{X}_I, \boldsymbol{X}_k)$, for $k=1,...,n_{val.train}$ in the training set, where $d(\boldsymbol{X}_I, \boldsymbol{X}_k)$ is the similarity metric between the index patient and the $k^{th}$ patient. 
            \item Use the $M$ obtained in step 4(b) to form the subpopulation.
            \item Fit the model of interest on the subpopulation. 
            \item Measure the performance of the model using, for example, the discrimination and calibration measures discussed in Section 2.3.
        \end{enumerate} 
        \item To quantify the uncertainty of the performance measures, fit bias-corrected and accelerated (BCa) bootstrap confidence intervals for each of the performance measures using the $B$ bootstrap estimates.
    \item Repeat Step 1-6 $Z$ times, ie. repeat the process with $Z$ different validation holdout samples (and, thus, TrTe samples) to view consistency of the tuning of $M$ (step (3) above) as well as the model performance measures in different hold-out validation samples. We suggest $Z$ to be approximately ten.
\end{enumerate}

Although seemingly very similar, there are two important distinctions between this algorithm and the one in \cite{krikella2025personalized}. Firstly, we updated the previous loss function to be more general, shown in Equation (\ref{eq: generalLoss}), in the sense that we do not specify which measures of discrimination and calibration to be included in the function, they are user-specified. Thus, in this algorithm, we have included Step 2 which involves choosing measures of discrimination and calibration to include in the defined loss function. Secondly, this algorithm contains specific instructions on how to define the grid of values on which to tune $M$ with the intention of lowering the computation time of running this algorithm. We specifically motivate and state our recommendations for these bounds in Section 3. Also, if the user wishes, $\alpha$ could be tuned as an additional step in the above algorithm.

\section{Simulation Studies}
\label{s:simulation2}

We conducted two sets of simulation studies using R version 4.4.1 \cite{Rprogramming}. The first involves investigating the relationship between the size of subpopulation and six different performance measures under various outcome models. The goal of this simulation study was to determine if the relationships found in \cite{krikella2025personalized} are robust against various true underlying outcome models. In the second simulation study, we demonstrate our previously published algorithm but with bounds put on the grid of values on which $M$ is tuned to lessen the computational burden, and using the updated loss function with the lack of spread as the discrimination measure and the ICI as the calibration measure.   

\subsection{Setup} 
\label{ss:Setup}

We considered a total sample size of $n=10 000$ and $40$ predictors. 

We replicated a scenario in which we have two clusters of patients of size $n_1 = 5000$ and $n_2 = 5000$, each cluster differing in the distributions of their respective predictors, as well as the strength of their association to the binary outcome of interest. Both patient clusters have the same predictors relating to the outcome, but the coefficients may differ. In the first cluster of patients, of the $40$ predictors generated, $20$ predictors were continuous with eight predictors simulated from a $N(30, 4)$ distribution, five predictors simulated from a $N(5, 1)$ distribution and seven predictors simulated from a $N(89, 29)$ distribution. We then simulated seven predictors from a $Bernoulli(0.3)$ distribution, nine predictors from a $Bernoulli(0.5)$ distribution and four predictors from a $Bernoulli(0.7)$ distribution. In the second cluster of patients, $20$ predictors were continuous with eight predictors simulated from a $N(20, 8)$ distribution, five predictors simulated from a $N(15, 8)$ distribution and seven predictors simulated from a $N(50, 9)$ distribution. We then simulated seven predictors from a $Bernoulli(0.5)$ distribution, nine predictors from a $Bernoulli(0.15)$ distribution and four predictors from a $Bernoulli(0.45)$ distribution. These two clusters of patients were included in one dataset of size $n_1 + n_2 = n = 10 000$. We standardized each predictor within this dataset so that all values of each predictor were between -1 and 1. This specific standardization was employed to be naturally incorporated into the cosine similarity metric calculation.

We defined 12 different true outcome models using the dataset described above. We simulated six outcomes from a linear model which differed by the number of predictors associated with the outcome, which could be either 10/40 or 20/40, or by prevalence of the outcome, which were low($\sim$0.05), moderate ($\sim$0.15) or balanced ($\sim$0.5). We also simulated six outcomes from a non-linear model which differed by the number of predictors associated with the outcome, or by prevalence, using the same specifications described above.

We provide examples of two of these outcome models in Equations (\ref{eq: exampleModelLinear}) and (\ref{eq: exampleModelNonLinear}). Specifically, Equation (\ref{eq: exampleModelLinear}) is a linear model which results in an outcome with low prevalence, and has ten predictors associated with the outcome, and Equation (\ref{eq: exampleModelNonLinear}) is a non-linear model which results in an outcome with moderate prevalence, and has ten predictors associated with the outcome. In this model, $z_1$ and $z_2$ are the predictors of the first and second cluster of patients, respectively. Each cluster of patients has the same predictors associated with the outcome, but these relationships generally differ in direction and/or strength of the association. 

\begin{align}
\label{eq: exampleModelLinear}
    z_1 & = &  -3 -3x_1 + x_2 + 0.5x_9 + x_{16} + x_{19} + x_{22} + x_{30} \notag \\ 
        &   & - x_{34} - 4x_{38} - 5x_{39} + \epsilon \notag  \\ 
    z_2 & = & -3.5 + 3x_1 + x_2 - x_9 + 2x_{16} + x_{19} + x_{22} + 4x_{30} \notag \\ 
        &   & + 5x_{34} + 3x_{38} + 3x_{39} + \epsilon 
\end{align}

\begin{align}
\label{eq: exampleModelNonLinear}
    z_1 & = &  -2x_1x_{39} - 4\exp({x_6x_{30}}) - 2\sin({x_9x_{16}})  \notag \\
    & & +0.01(x_{19})^3 +3x_{22}x_{34} + 2x_{38} + \epsilon \notag \\ 
    z_2 & = & x_1x_{39} - \exp({x_{19}x_{38}}) - 2\cos({x_9x_{16}}) \notag \\
    & & +0.01(x_{6})^2 - 3x_{22}x_{34} + 3x_{30} + \epsilon
\end{align}

After the outcome was simulated from one of the 12 different models, both $z_i$, $i = 1,2$ were passed through an inverse logit function to obtain two probabilities, $\pi_i$, $i=1,2$, as shown in Equation (\ref{eq: inverseLogit2}). These probabilities were then used to simulate Bernoulli outcomes, $y_{ki}$, where $Y_{ki} \sim Bernoulli(\pi_i)$. Note that the simulated outcome of each index patient in the testing dataset, $y_{Ii}$, was generated in the same way.

\begin{equation}
\label{eq: inverseLogit2}
     \pi_i = 1/(1+\exp(-z_i)).
\end{equation}  

The final dataset includes the $40$ predictors and the binary outcome of all $n = 10 000$ patients. We stress again that, as we considered 12 different outcome models, we ultimately simulated 12 different datasets. 

In all simulation studies, we simulate 10 randomly generated datasets and, within each dataset, we conducted 20-repeated 10-fold cross validation. The reason we only ran 10 simulated datasets is due to the considerable computational expense of each simulation run. This is a point we further address in Section 5.

In the validation step, we calculated a BCa bootstrap confidence interval for each performance measure, where we simulated $B=1000$ bootstrap samples. 

\subsection{Relationships between the Size of Subpopulation and the Performance Measures} 
\label{ss: sim1}

In this section, we display the results from two of the 12 datasets. Dataset 1 has a linear true outcome model with 20 predictors associated with the outcome and a resulting prevalence that is approximately 0.15. Dataset 2 has a non-linear true outcome model, also with 20 predictors associated with the outcome and a resulting prevalence that is lower, at approximately 0.05. These descriptions are displayed in Table 1, and we will refer back to these datasets throughout this section. 

\begin{table}[h!]
\caption{Characteristics of two of the twelve datasets simulated.}
\centering
\begin{tabular}{|l|l|l|}
\hline
                                                                                       & \textbf{Dataset 1} & \textbf{Dataset 2}  \\ \hline
\textbf{Linearity of outcome}                                                                  & Linear    & Non-Linear \\ \hline
\begin{tabular}[c]{@{}l@{}}\textbf{Number of predictors}\\ \textbf{associated with outcome}\end{tabular} & 20        & 20         \\ \hline
\textbf{Outcome prevalence}                                                                     & 0.15      & 0.05       \\ \hline
\end{tabular}
\label{table: datasets}
\end{table}

We analyze the relationships between $M$ and six performance measures (AUROC, AUPRC, lack of spread, CITL, calibration slope and ICI) for Dataset 1 from Table 1, shown in Figure 1. The relationship between $M$ and the AUROC and AUPRC, as seen in Figure 1(a) and (b), is negatively proportional after $M = 3000$, ie. the discrimination improves up to a point, where it then begins to deteriorate. Under the lack of spread measure, displayed in Figure 1(c), the relationship is negatively proportional, even for the lowest values of $M$. As $M$ increases, the discrimination deteriorates, meaning that a low value of $M$ will result in a PPM with relatively better discriminative ability than a higher value of $M$. Figure 1(d)-(f) displays the relationship between $M$ and calibration using the three increasingly strict measures of calibration: the CITL, calibration slope and the ICI. The calibration slope and the ICI both have a concave quadratic relationship with $M$, with both low and high value of $M$ resulting in models with very good calibration. Under calibration slope, the best performing model is when $M = 1250$, which is the second best performing model under the ICI; the best, only by a difference of 0.003, being under $M=9000$ for the latter measure of calibration. The CITL has a convex relationship with $M$, and is minimized at $M=4000$. We do note that the range of CITL values, which has differences in the thousandths, specifically 0.003, is much smaller compared to the calibration slope and the ICI, which have a ranges of 0.631 and 0.017, respectively. 

 \begin{figure*}[h!]
    \centering
        \subfloat[AUROC]{%
            \includegraphics[width=.35\linewidth]{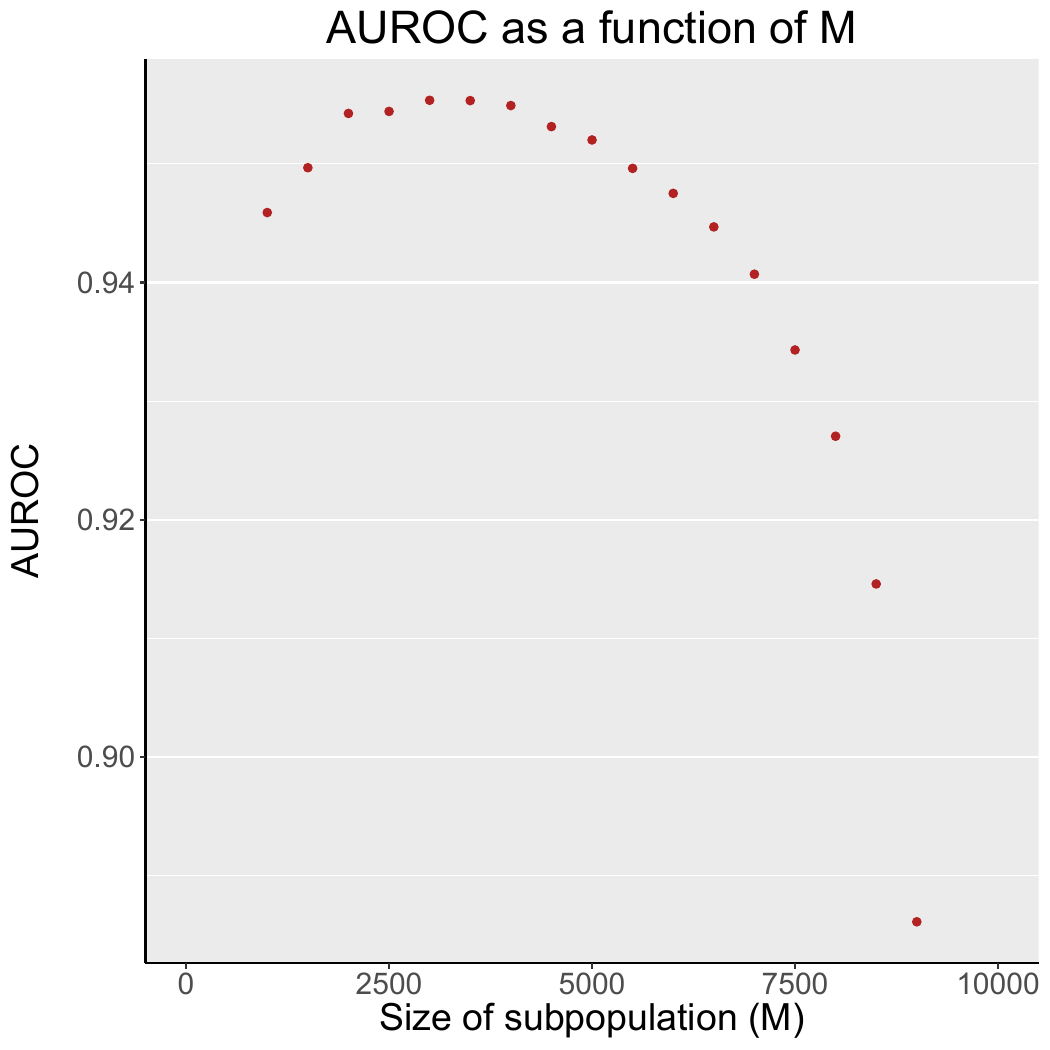}%
        } 
        \subfloat[AUPRC]{ 
            \includegraphics[width=.35\linewidth]{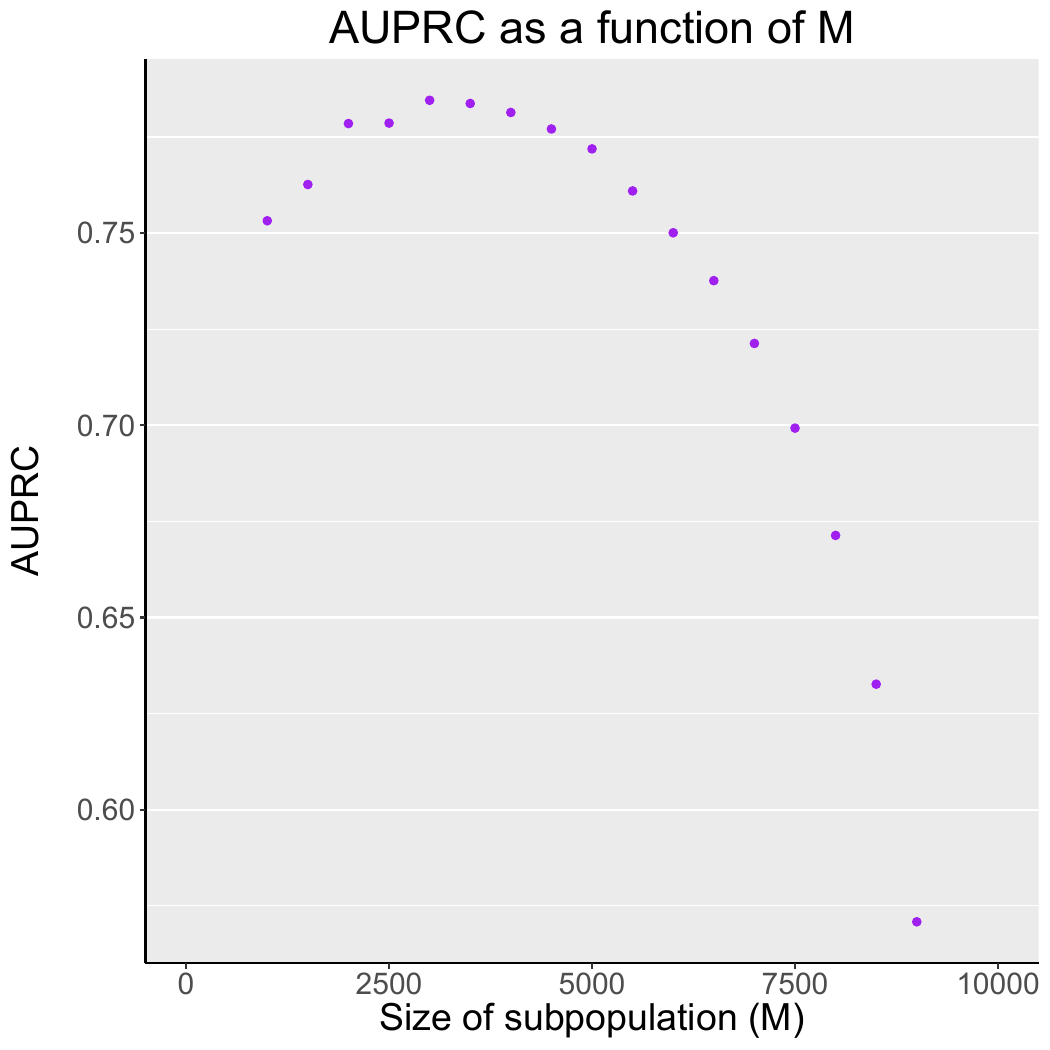}%
        } 
        \subfloat[Lack of Spread]{%
            \includegraphics[width=.35\linewidth]{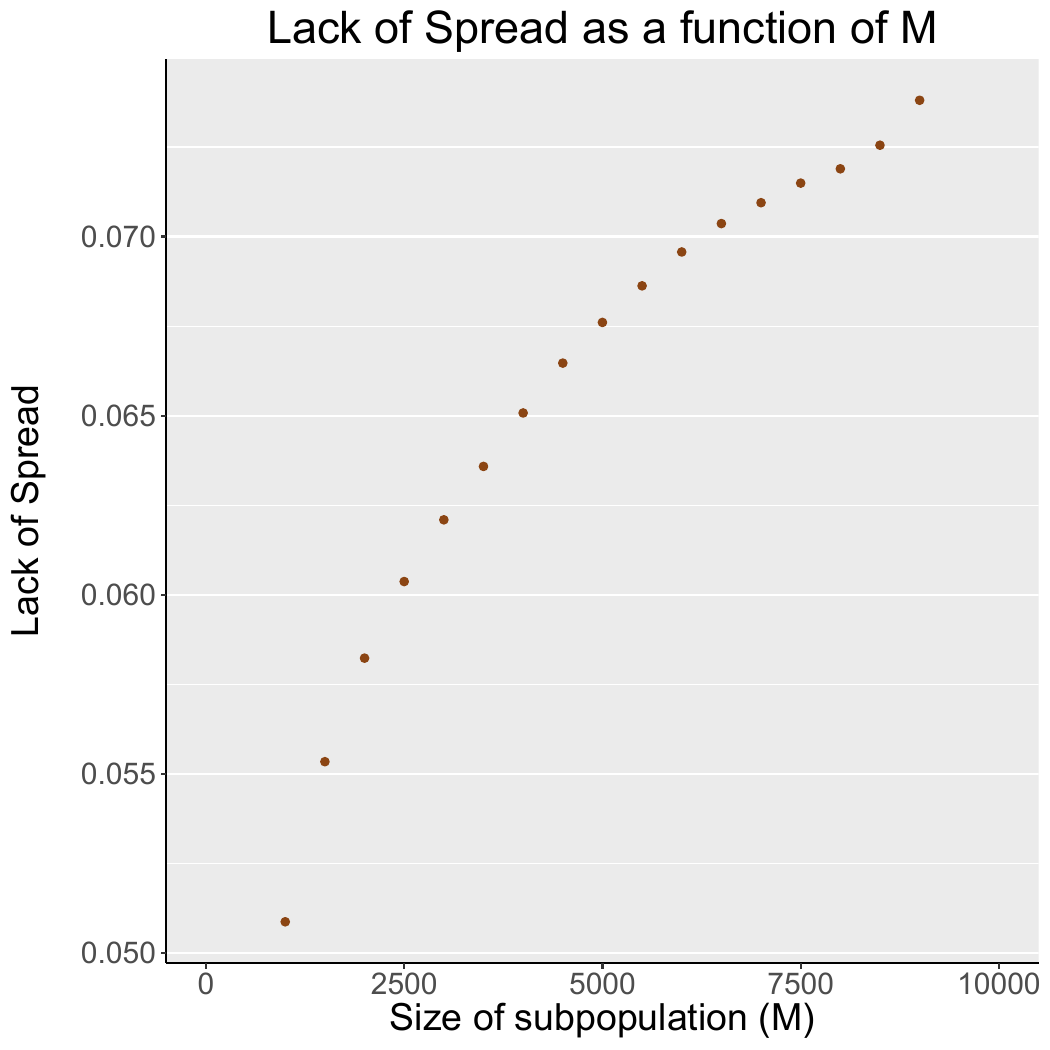}%
        } \\
         \vspace{1em} 
        \subfloat[CITL]{%
            \includegraphics[width=.35\linewidth]{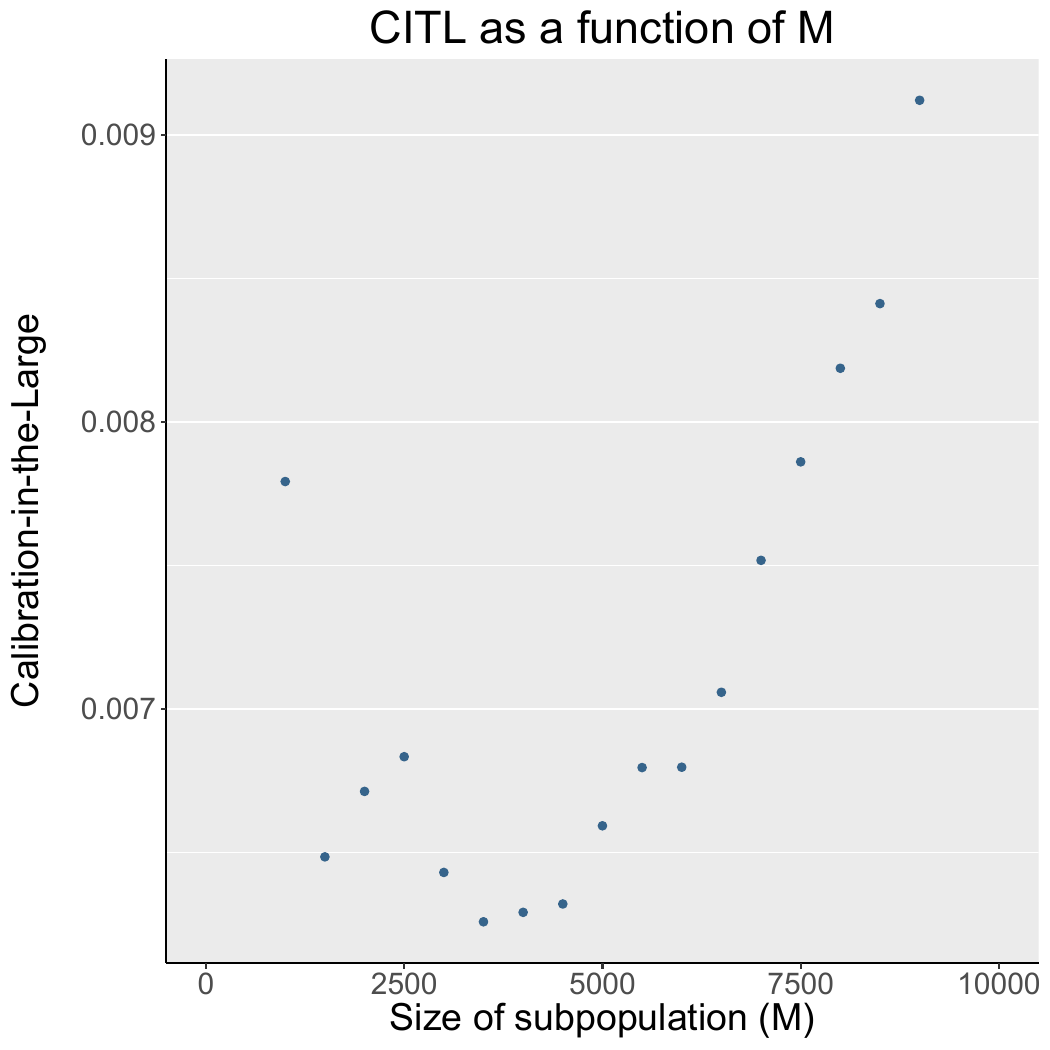}%
        } 
        \subfloat[Calibration Slope]{%
            \includegraphics[width=.35\linewidth]{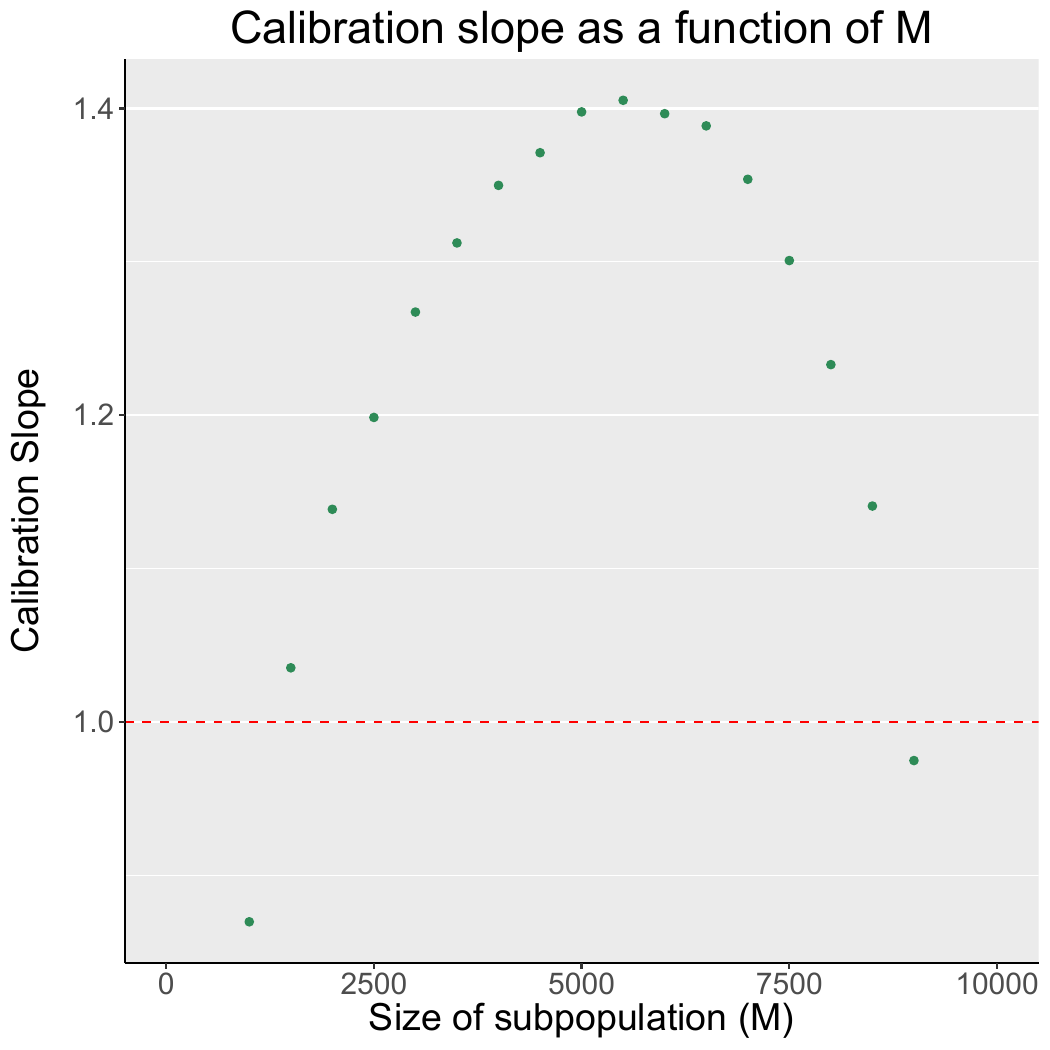}%
        }
        \subfloat[ICI]{%
            \includegraphics[width=.36\linewidth]{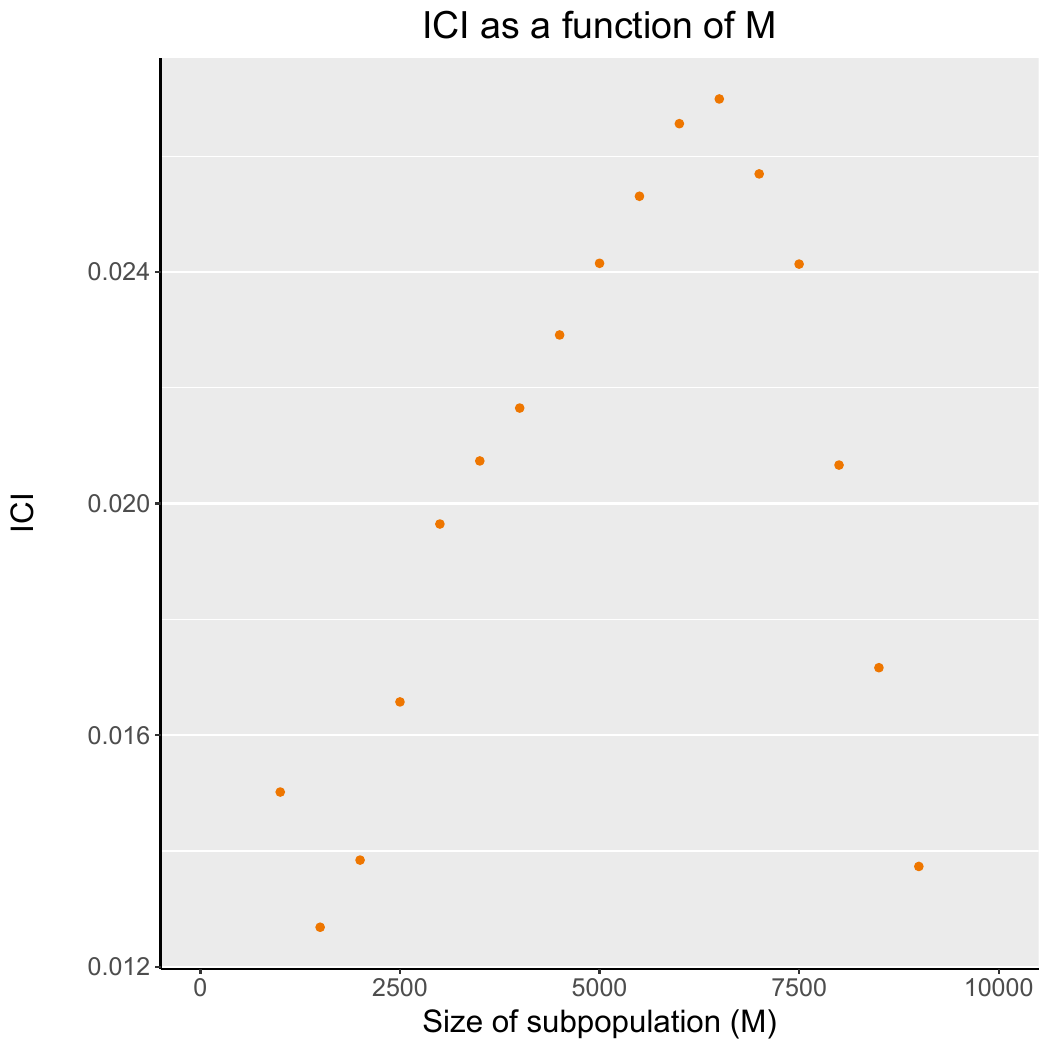}%
        } 
        \caption{Performance measures as a function of $M$ for Dataset 1 from Table \ref{table: datasets}.} 
        \label{NonLinLowTwenty}
    \end{figure*}

The relationship between $M$ and the same six performance measures for Dataset 2 from Table 1 is shown in Figure 2. As expected, the AUROC and the AUPRC improve up to a point, specifically at $M=2500$, where they then begin to deteriorate as $M$ increases. We note, however, that it might be argued that the AUROC is not the best measure of discrimination to consider under an outcome proportion of near 0.05. The relationship between lack of spread and $M$, shown in Figure 2(c), is also generally negatively proportional, however, at $M=3500$, the discrimination under this measure begins to improve slightly before deteriorating again. Also, there is an outlier at $M=9000$ that indicates improvement in the lack of spread from the previous $M$ value of $M=8500$. We would like to note that out of the 12 datasets considered, this case is the only one which did not result in a true negative proportional relationship between $M$ and the lack of spread. The complexity of the outcome model (ie. non-linear, low prevalence and 20 predictors associated with the outcome) may have had an effect on this relationship. 

Both the CITL and the ICI, as shown in Figure 2(d) and (f), are optimal under relatively larger values of $M$, the former being minimized at $M=7250$ and the latter at $M=6500$. The relationship between $M$ and the calibration slope is, once again, concave, seen in Figure 2(e), and is optimal, or a negligible distance away from optimal, at both $M=2250$ and $M=8500$. Similar to what was seen in the previous case, the second best choice of $M$ under the ICI is the same value which optimizes the calibration slope: $M=2250$. All three measures are in agreement in terms of the worst calibrated model, which in the case of this dataset, is the model which is fit on the smallest size of subpopulation considered: $M = 1000$. The difference between the best calibrated PPM and worst calibrated PPM, in terms of the CITL, is 0.014, which is a much wider range compared to the range of CITL values in the previous dataset. The calibration slope also has a much wider range compared to what was seen in Figure 1, where here, the difference between the best and worst calibrated model is 0.989. The ICI has a similar range to the previous example, valued at 0.016. In this simulation study, higher values of $M$ perform better than lower values of $M$ in terms of calibration under all three measures, however, a lower value of $M$ still results in a model with good weak and moderate calibration.

We would like to point out that the relationships observed in the above two cases are not unique to the specific characteristics of the datasets considered. For instance, Dataset 2 has an outcome that is generated from a non-linear model, and in Figure 2, we see that the optimal model under the ICI is one which is trained on approximately 65\% of the training dataset. However, there were other datasets we considered which had outcomes generated from non-linear models and resulted in a low optimal $M$ proportion under the ICI. We cannot conclude that any of these results are due to the specific characteristics we varied, however, we can make some overall conclusions that we discuss below.

\begin{figure*}[h!]
    \centering
        \subfloat[AUROC]{%
            \includegraphics[width=.35\linewidth]{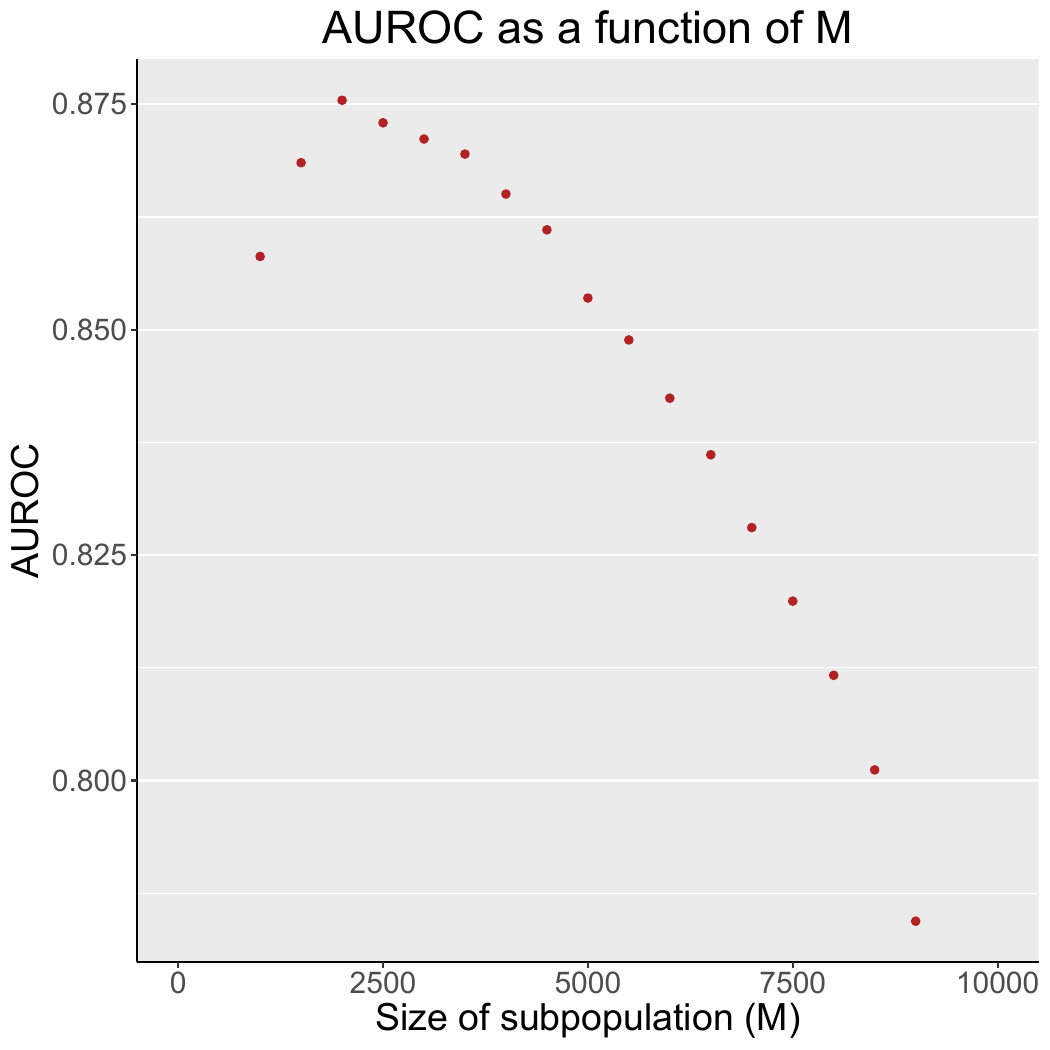}%
        } 
        \subfloat[AUPRC]{ 
            \includegraphics[width=.35\linewidth]{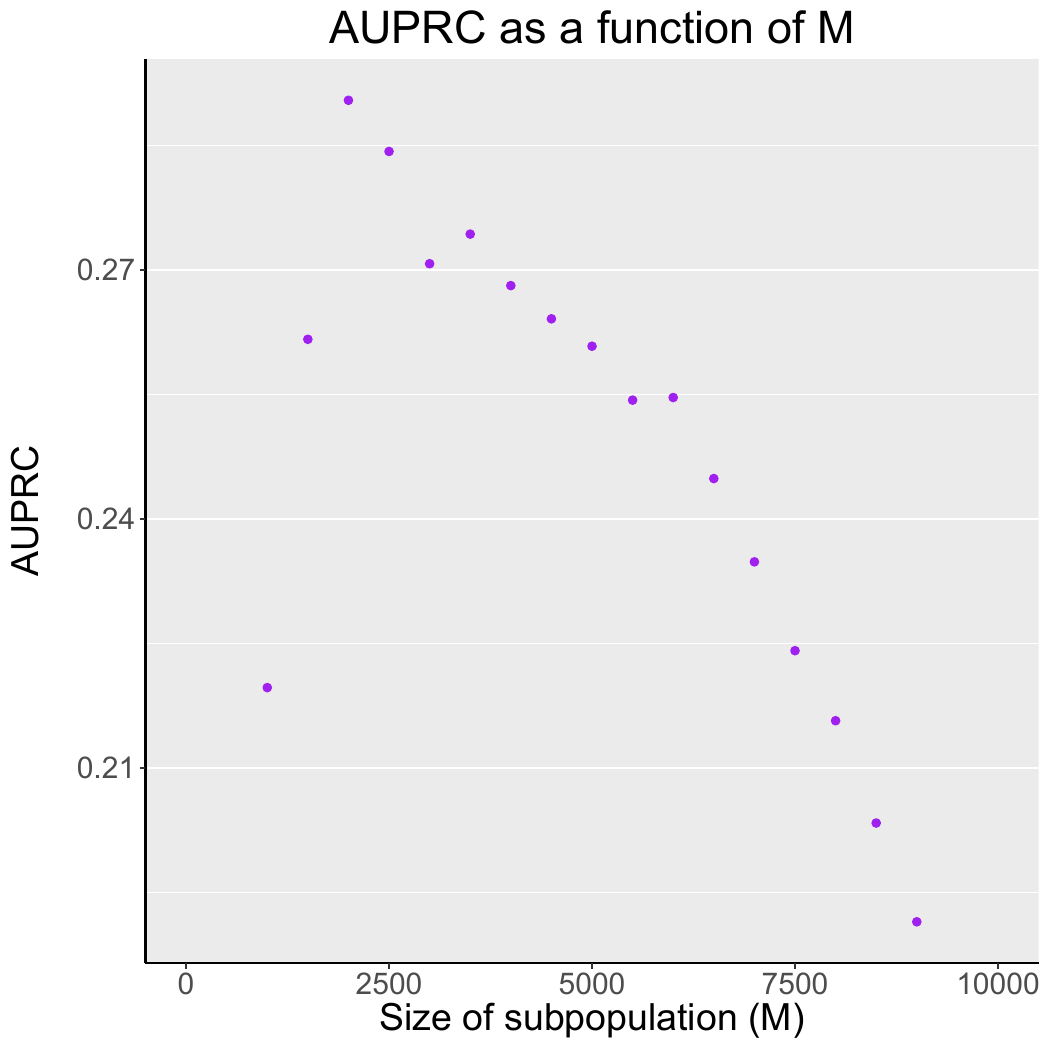}%
        } 
        \subfloat[Lack of Spread]{%
            \includegraphics[width=.35\linewidth]{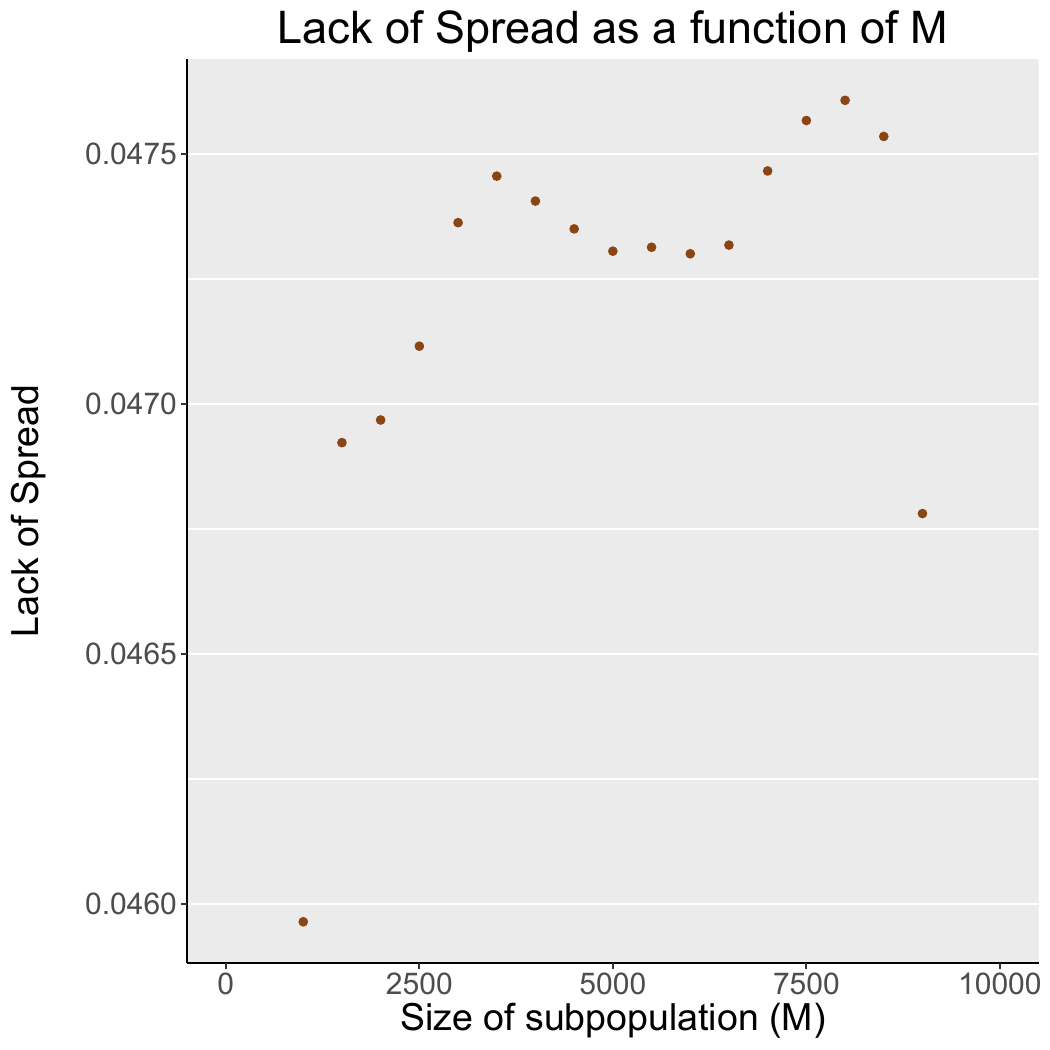}%
        } \\
        \vspace{1em}  
        \subfloat[CITL]{%
            \includegraphics[width=.35\linewidth]{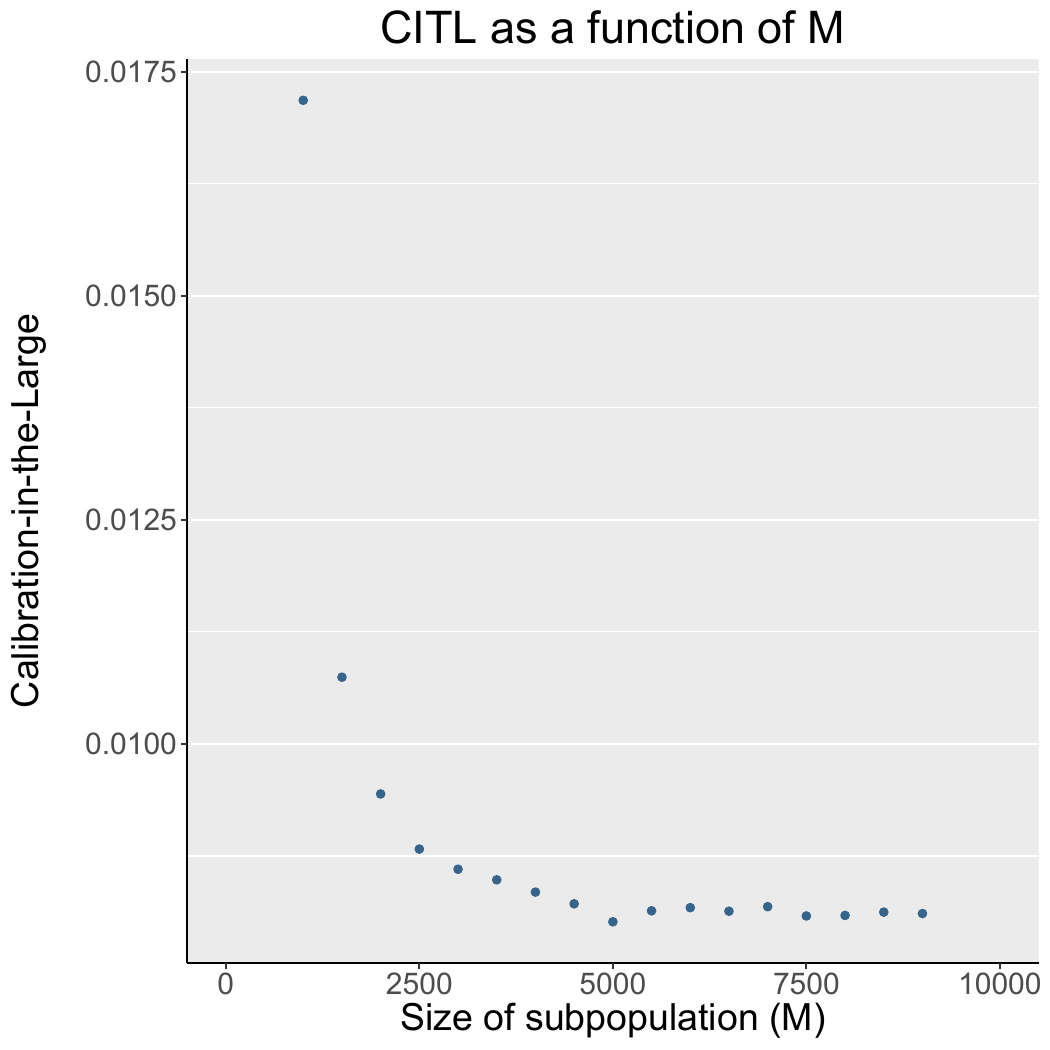}%
        } 
        \subfloat[Calibration Slope]{%
            \includegraphics[width=.35\linewidth]{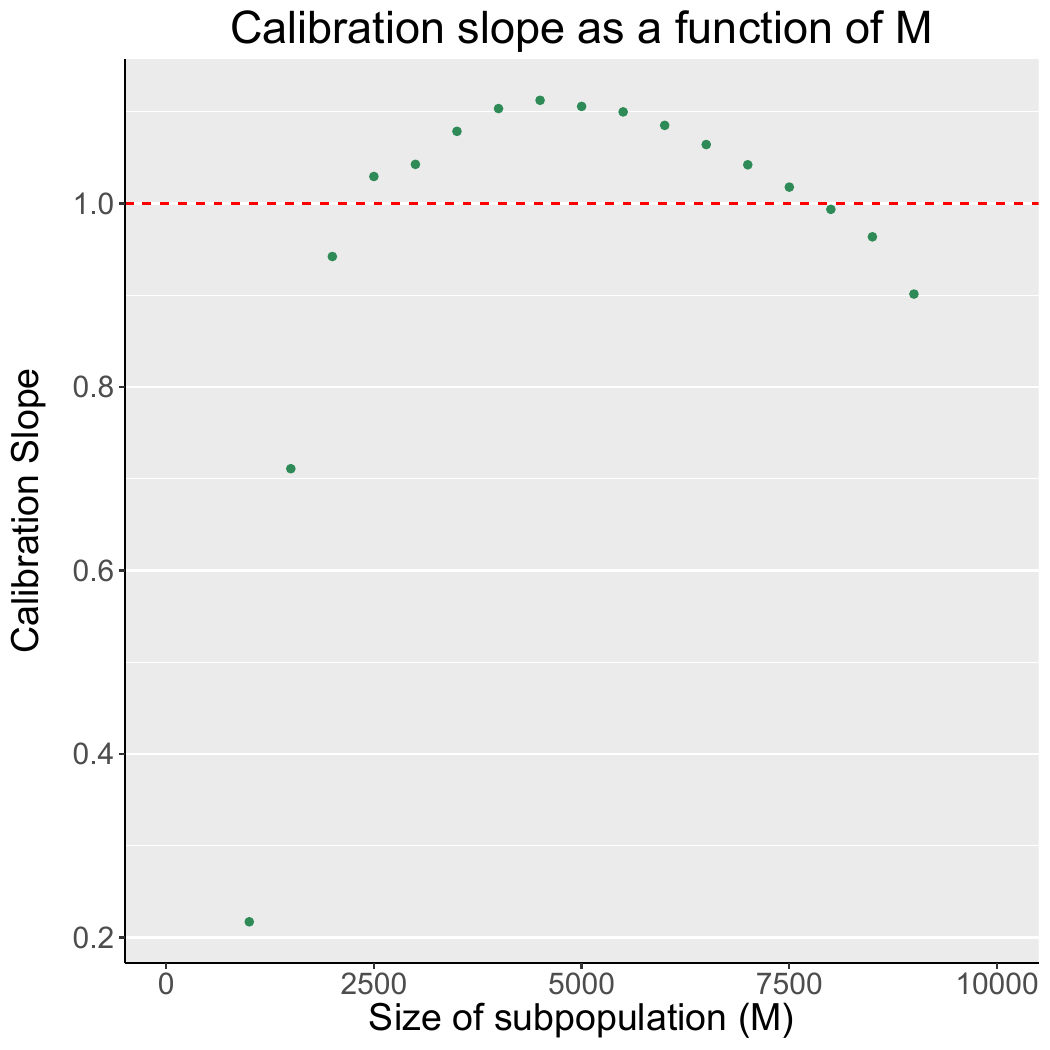}%
        }
        \subfloat[ICI]{%
            \includegraphics[width=.36\linewidth]{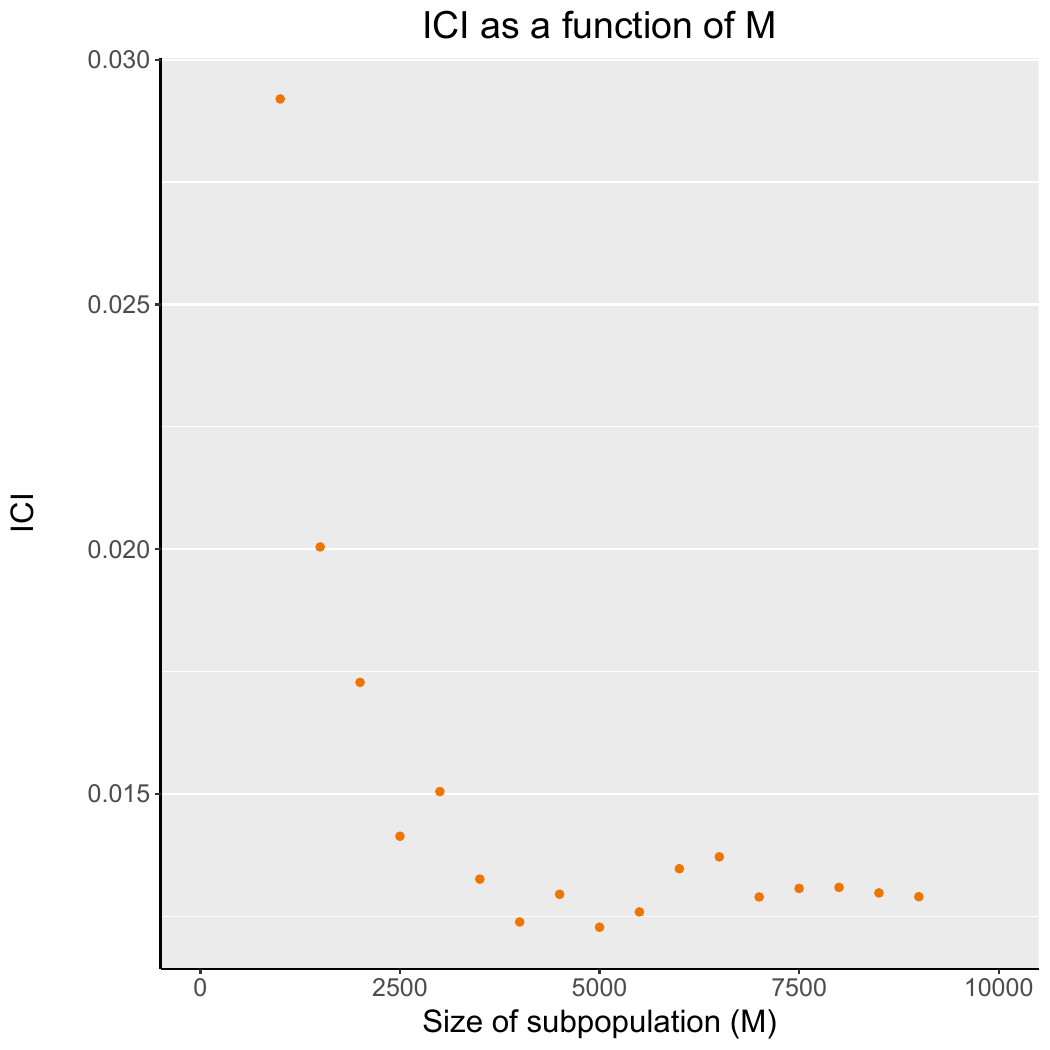}%
        } 
        \caption{Performance measures as a function of $M$ for Dataset 2 from Table \ref{table: datasets}.} 
        \label{NonLinLowTwenty}
    \end{figure*}

Under all 12 simulation studies, the relationship between $M$ and discrimination was negatively proportional, aside from the lowest values of $M$ when using the AUROC or the AUPRC as measure of performance. In terms of calibration, the relationships between $M$ and calibration differed under different measures, as well as under different datasets. In many cases, a low $M$ was good, if not the best, under the ICI. The calibration slope, in most cases, had a concave relationship with $M$, resulting in relatively better weak calibration in models fit on both small and large $M$ values. Due to this, there was usually agreement between the optimal $M$ under the calibration slope and the ICI. The optimal $M$ under the CITL was more sensitive to changes in dataset, with no clear pattern amongst the 12 cases, but in many cases, it was not in agreement with the optimal $M$ value found under the slope or ICI. As measures of moderate calibration ensure that models have good mean, weak and moderate calibration, we emphasize the recommendation by \cite{van2016calibration} and state that, in the general loss function shown in Equation (\ref{eq: generalLoss}), although there is flexibility in which calibration measure to use, we encourage the use of a measure of moderate calibration, such as the ICI. 

Based on these results, we recommend that when tuning $M$, the lower bound on the grid of values should be $0.2n_{train}$, where $n_{train}$ is the size of the training set (in this case, $n_{train} = 0.9*n_{TrTe}$ to account for the 10-fold cross validation), and the upper bound on the grid of values to be either $0.5n_{train}$ or $0.7n_{train}$, depending on the interests of the study. We recommend the lower bound be 20\% of the training data as we see that this proportion gives at the very least, good, if not the best, discriminative ability under the AUROC and AUPRC while still ensuring that the sample size in the validation step is sufficient. The upper bound on this grid of values will depend on the $\alpha$ value chosen, specifically, how much of an interest there is in the calibration of the model. For $\alpha = 0.5$ or lower, thus if there is equal interest in both discrimination and calibration, or even a greater interest in discrimination,  we suggest an upper bound of 50\%, which greatly reduces the values on which to tune $M$ while still capturing values which are known to include the optimal size under discrimination and, possibly, the optimal size under calibration as well. For larger values of $\alpha$, ie. values of $\alpha$ greater than 0.5, when there is greater interest in calibration, we recommended an upper bound of 70\% of the training data, which could capture the optimal value of $M$ in cases where that value is large in terms of calibration. We do not recommend a value higher than 70\% as an $M$ value greater than this results in a significant deterioration in the discriminative ability of the model.

\subsection{Relationship Between the Size of Subpopulation and the Mixture Term}

Now that we have a defined lower bound on the grid of values on which we tune $M$, we can consider values of $\alpha$ lower than 0.475 (ie. the lowest value of $\alpha$ considered in \cite{krikella2025personalized}) since the defined bound ensures there will be a sufficient sample size for the validation step. Thus, in this simulation study, we consider the following values of $\alpha$: $0.5$, $0.5 \pm 0.1$, $0.5 \pm 0.25$, $0.5 \pm 0.4$. We choose 0.5 as the baseline since this weight will put equal value on both the discrimination and calibration terms in the loss function. 

We begin by considering Dataset 1 from Table 1. Recall in Figure 1, a low value $M$ is associated with good discrimination as well as good weak and moderate calibration, but a value of $M$ that is approximately 50\% of the whole training data leads to a model with good mean calibration. We applied our proposed algorithm from Section 2.4, and we defined the measure of calibration in the loss function to be the ICI and the measure of discrimination to be the lack of spread to, ultimately, obtain the loss function shown in Equation (\ref{eq: newLoss}). Figure 3(a) shows that the resulting optimal $M$ under the seven values of $\alpha$ considered is low. This plot shows the relationship between $M$ and $\alpha$ using five different randomly generated datasets, and is in line with the information displayed in Figure 1. In Figure 1, the optimal value of $M$, of one seed considered, under the ICI was $M = 9000$, but the difference between this optimal model and the second optimal model was very small, thus in this simulation study where we conducted 20 repeated 10-fold cross validation, across 10 different seeds, the more robust result shows that the smaller value of $M$ was, indeed, the optimal value. For all 10 randomly generated datasets, the optimal $M$ proportion is approximately 0.22 under all values of $\alpha$, though in the plot we only display five out of 10 seeds.

Next, we considered Dataset 2 described in Table 1. As shown in Figure 2, high values of $M$ are associated with better mean, weak, and moderate calibration, while low values of $M$ are associated with better discrimination. 
In Figure 3(b), it is shown that as we increase $\alpha$, the optimal $M$ proportion also increases, under all 10 randomly generated datasets, though only five out of 10 are shown in this plot. For higher values of $\alpha$, there is more emphasis on the moderate calibration term in our loss function, thus these results coincide with the results seen in Figure 2: a higher value of $M$ is associated with good moderate calibration. For low values of $\alpha$, the optimal $M$ proportion is between 0.2 and 0.4. For high values of alpha, the optimal $M$ proportion is between 0.6 and 0.7. Unlike in Figure 3(a), there is more variability in the optimal $M$ proportions, especially for values of $\alpha$ that are in the middle of the range (ie. 0.4, 0.5, 0.6). However, the results are still consistent with what we expected as there is still the increasing relationship between $M$ and $\alpha$, regardless of the outliers.

 \begin{figure*}[h!]
    \centering
        \subfloat[Dataset 1]{%
            \includegraphics[width=.5\linewidth]{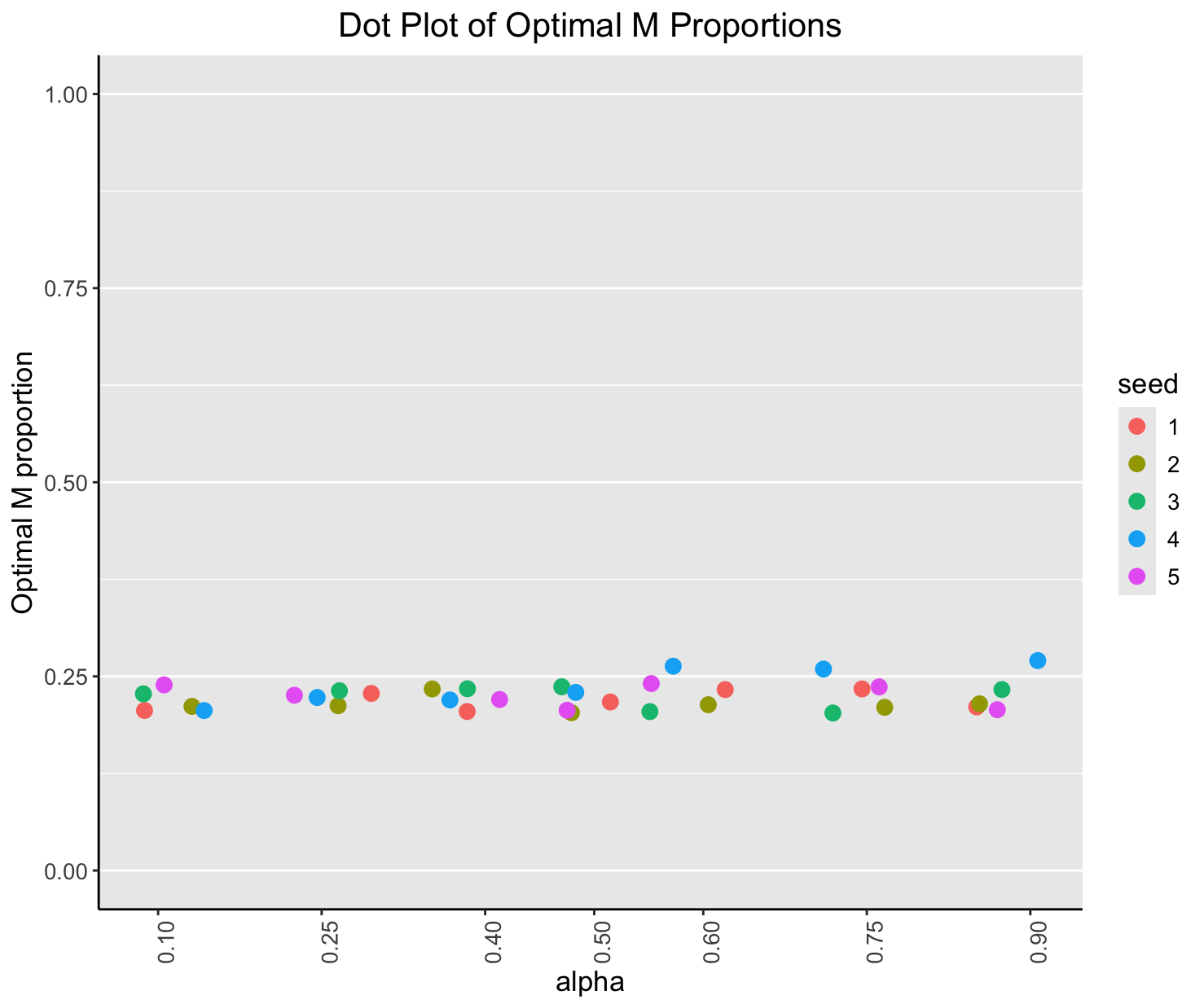}%
        } 
        \subfloat[Dataset 2]{ 
            \includegraphics[width=.5\linewidth]{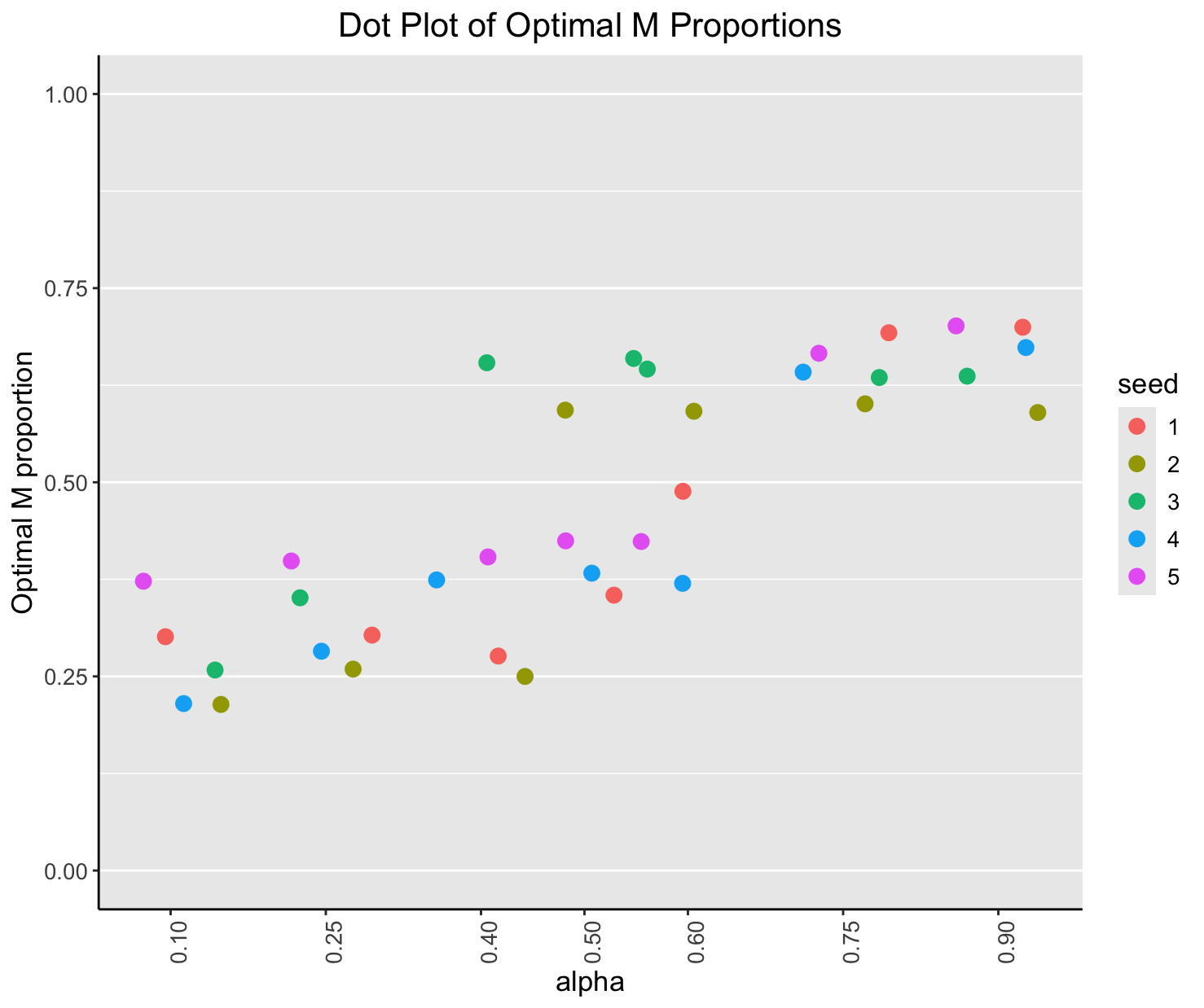}%
        } \\
        \caption{Performance measures as a function of $M$ for Datasets 1 and 2 from Table \ref{table: datasets}.} 
        \label{fig: optimalLin}
    \end{figure*}

\section{Data Analysis} 
\label{s:data2}

We applied our proposed algorithm to a dataset from the eICU Database \cite{pollard2018eicu} to predict mortality of patients with diseases of the circulatory system.

This dataset only considers patients in their first ICU stay of their first hospital visit and had a disease of the circulatory system, identified by their ICD9 code. These patients have survived in the ICU for at least 24 hours, but their stay did not exceed 14 days. The minimum of 24 hours was chosen to ensure that we could have the necessary data obtained to calculate the useful APACHE (IVa) score, which is an improvement built upon Apache IV \cite{zimmerman2006acute}. The maximum of 14 days was chosen as these patients were outliers, with only 2\% of unique patients having ICU stays lasting more than 14 days. However, putting these bounds still gave us sufficient sample size to demonstrate our proposed methodology on this dataset. Any patient with missing values in any of the predictors, as well as any patient with no vital sign data collected in the first 24 hours of their ICU stay, was removed. This approach is not recommended in practice, but dealing with missing data was not the focus of the current manuscript. Our primary purpose is to demonstrate the proposed methodology in a real application dataset. In total, we have a cohort patient size of 23,928 patients. We used 80\% of this cohort as the TrTe set, where we then performed 10-fold cross validation. Thus, the TrTe set consisted of 19,143 patients, and once the TrTe set was split into distinct training and testing sets using 10-fold cross validation, the training set consisted of 17,229 patients and the testing set consisted of 1,914 patients.

For each patient, the following predictors were extracted:

\begin{itemize}
    \item Admission and demographic data: age, gender (male/female), and hospital discharge status (Expired/Alive). 
    \item Vital signs from the first 24 hours after the start time of ICU admission: oxygen saturation (sao2), respiratory rate (respiration), and heart rate.
    \item APACHE IVa score: a severity-of-illness score utilizing demographic information as well as medical information collected within the first 24 hours of a patient's ICU stay.
\end{itemize}

To identify the dominant modes of temporal variation of the vitals signs in the first 24-hours of a patient's stay in the ICU, we used functional principal components (FPCs) \cite{yao2005functional}. We extracted the first three FPCs for each of the vital signs, which cumulatively explained at least 95\% of the total variation in the trajectory of each given vital sign. Our resulting dataset thus consists of 13 predictors, including the first three FPC scores for each of the three vital signs.

When testing the algorithm, we considered $\alpha = \{0.1, 0.25, 0.4, 0.5, 0.6, 0.75, 0.9\}$. We conducted 20 repeated 10-fold CV in the training/testing step of the algorithm, then validated the results with a holdout sample. We fit a logistic regression model to the data as the outcome (hospital discharge status) is binary (i.e., expired or alive). As the prevalence of outcome in this dataset is approximately 0.107, we use the AUPRC as a measure of discrimination instead of the AUROC. We also include a measure of the lack of spread, and the three measures of calibration used in Section 3. 

We show the relationships between $M$ and the performance measures when fitting PPMs to this dataset in Figure 4. As expected, Figure 4(a) shows that the AUPRC of the model increases up to a point, but instead of decreasing as $M$ increases, the value of the measure stays relatively constant after this point, but still decreases slightly. The optimal $M$ value under the AUPRC is $M = 9000$. Figure 4(b) shows the relationship between $M$ and lack of spread, and, in general, as the size of subpopulation increases, the lack of spread of the model deteriorates, with the optimal $M$ value being $M=4500$. Figures 4(c) and (d) show that the optimal $M$ under mean and weak calibration is very high, however we do note that the differences in the CITL between models fit using varying values of $M$ is very small. Further, similar to what was seen in the simulation study, there are two values of $M$ that result in optimal models under the calibration slope, with the first optimal value being approximately 50\% of the entire training data, and the second being nearly the entire dataset. Under the ICI, the optimal $M$ is the smallest value considered, however, moderate calibration performs very well for high values of $M$ as well, as seen in Figure 4(e).  

\begin{figure*}[h!]
    \centering
        \subfloat[AUPRC]{%
            \includegraphics[width=.35\linewidth]{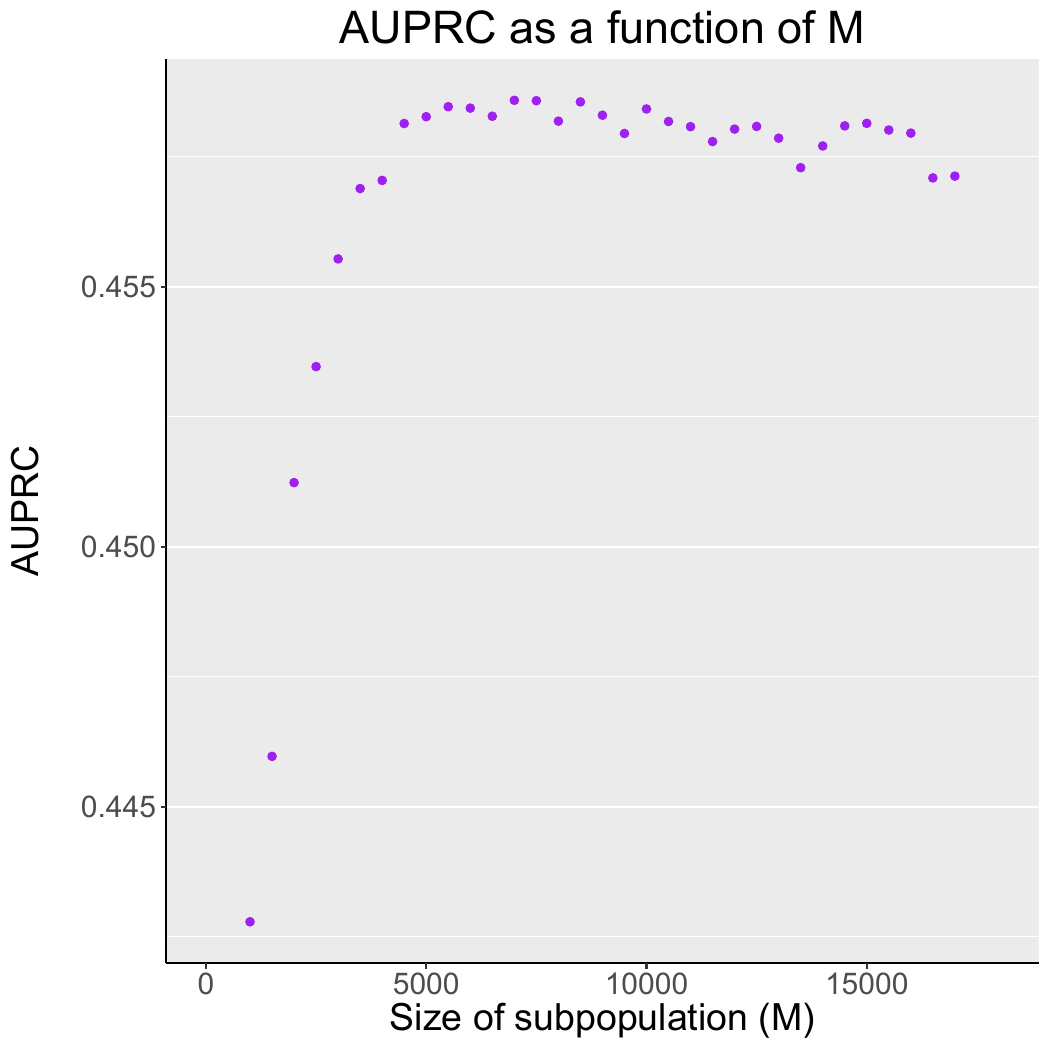}%
        } \subfloat[Spread]{%
            \includegraphics[width=.35\linewidth]{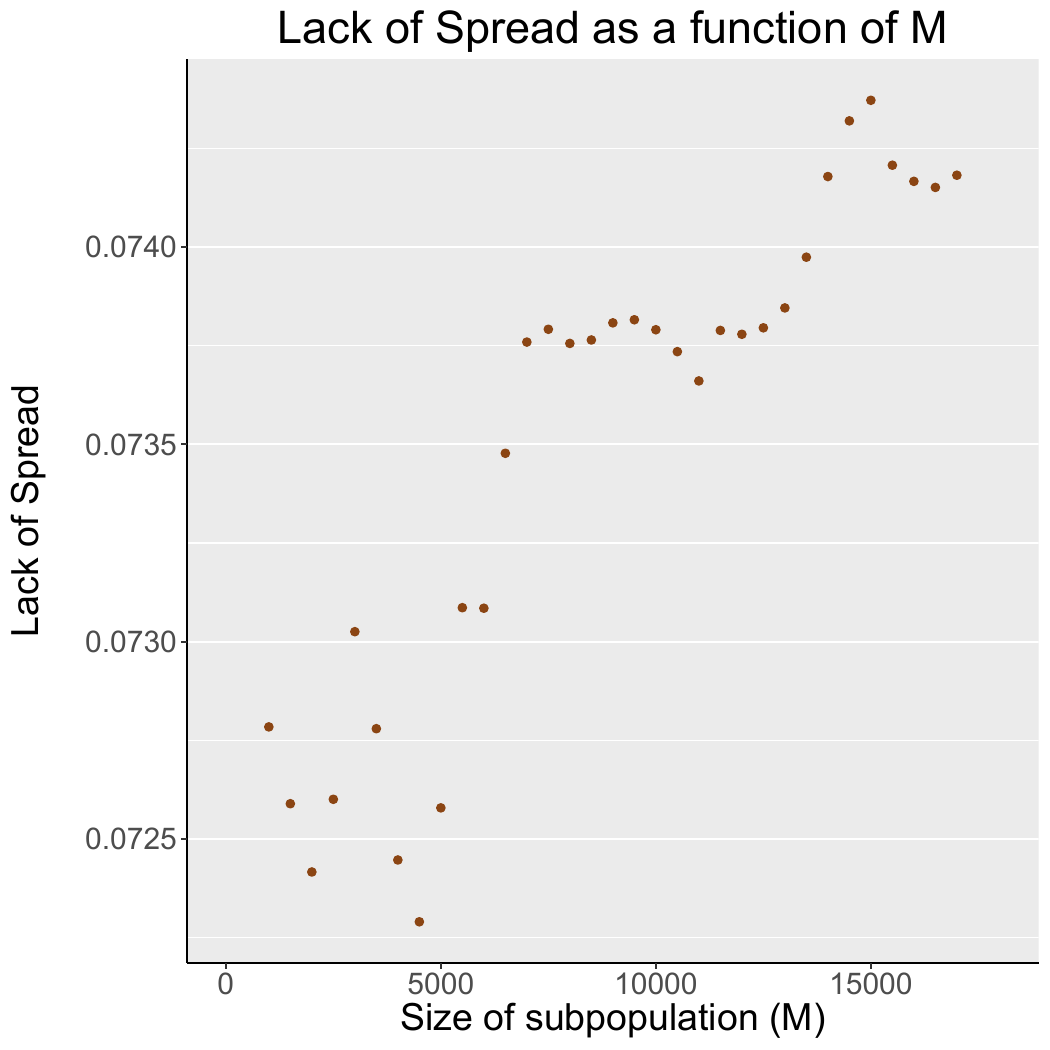}%
        }
        \subfloat[CITL]{%
            \includegraphics[width=.35\linewidth]{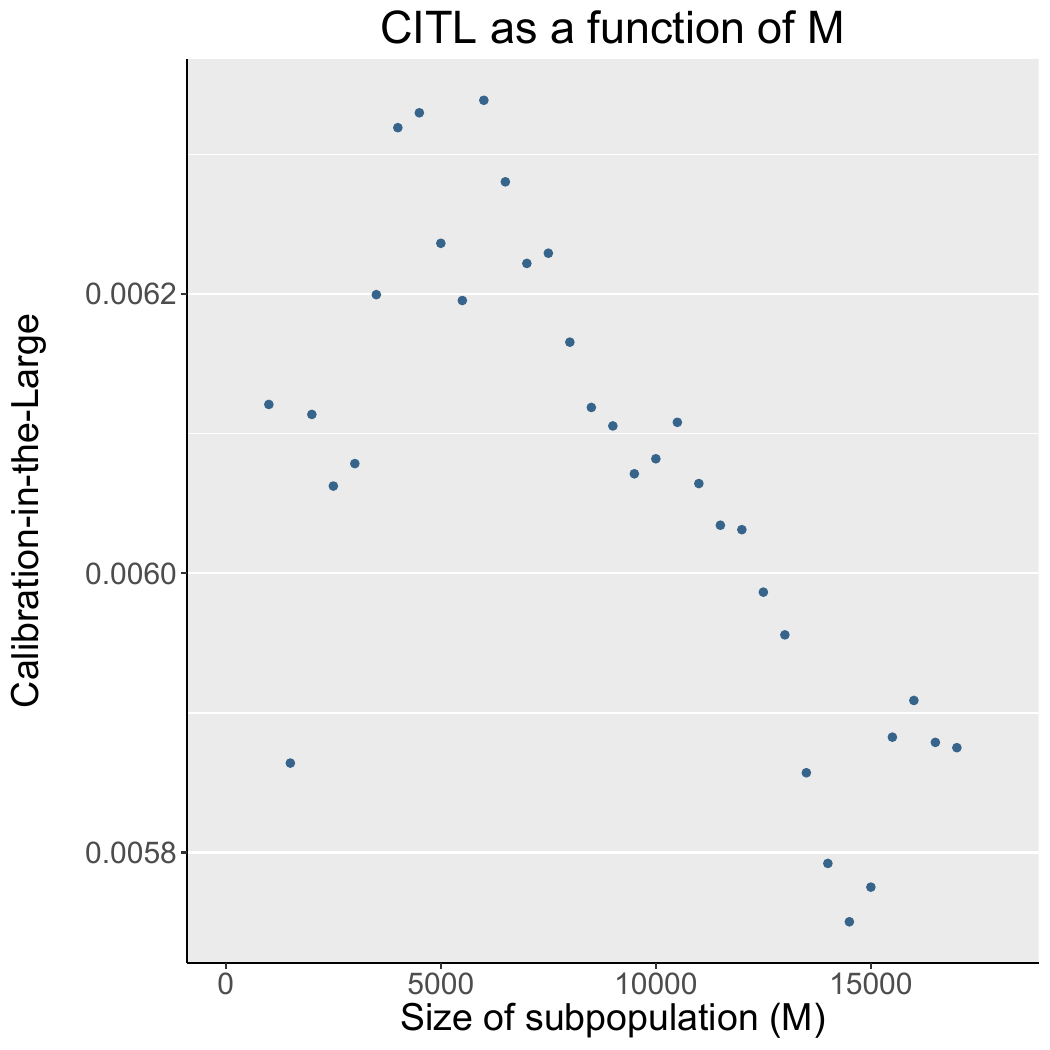}%
        } \\
        \vspace{1em}  
        \subfloat[Calibration Slope]{%
            \includegraphics[width=.35\linewidth]{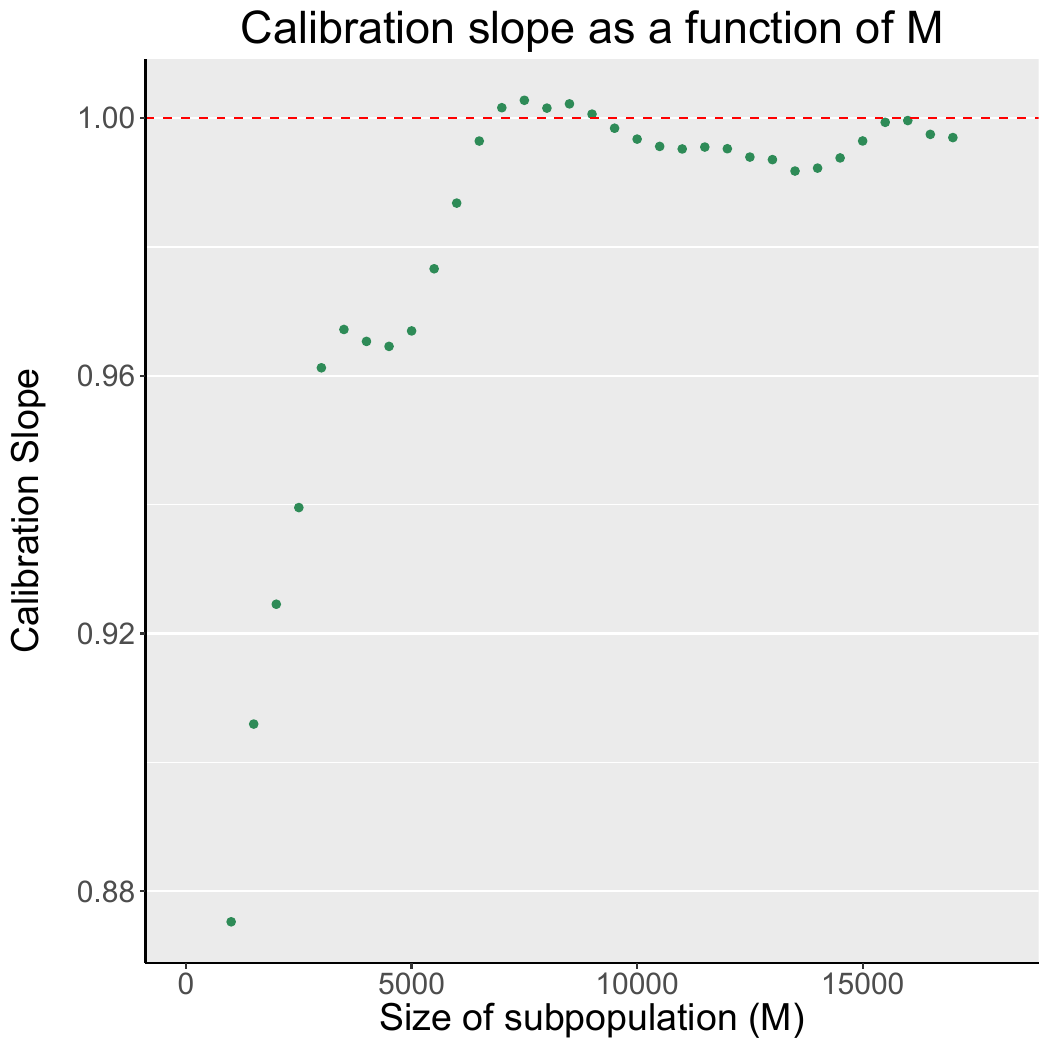}%
        }
        \subfloat[ICI]{%
            \includegraphics[width=.36\linewidth]{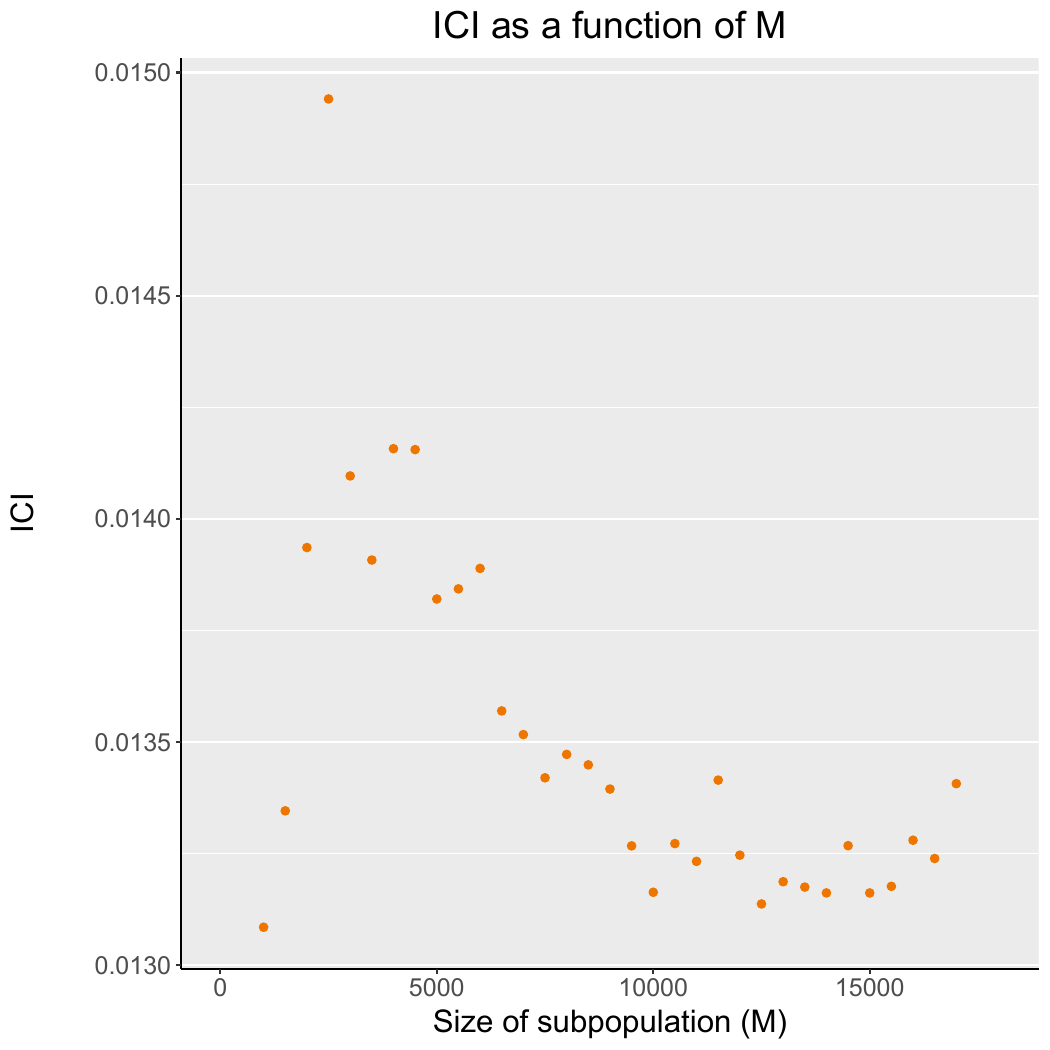}%
        } 
        \caption{Performance measures as a function of $M$ for the cardiac dataset from the eICU database.} 
        \label{CardiacRelationships}
    \end{figure*}

After investigating the relationships between $M$ and the performance measures, we then applied the proposed algorithm from Section 2.4 to the eICU data. We applied this algorithm twice, considering two different loss functions. The first is the same loss function used in the simulation studies, shown in Equation (\ref{eq: newLoss}), which consists of the ICI and the lack of spread measure, as the measures of calibration and discrimination, respectively. The second loss function we consider consists of the ICI as the calibration measure and the AUPRC as the discrimination measure, as shown below in Equation (\ref{eq: auprcLoss}). We approximate the AUPRC in the second term after the equal sign and define $P(c)$ as the precision at a cutoff $c$, $c =1,...,d$, and $\triangle r(c)$ as the change in recall that occurred between cutoff $c-1$ and cutoff $c$. We use (1-AUPRC) as the measure since higher values of the AUPRC correspond to better discriminative ability, and we wish to minimize $L^{**}$.

\begin{equation}
    \label{eq: auprcLoss}
    L^{**(M)} = \frac{\alpha}{n_{test}}\sum_{I=1}^{n_{test}}\left|\hat{p_I}^{(M)} - p_I^{(M)}\right|
    + (1-\alpha)\left[1 - \sum_{c=1}^dP(c)\triangle r(c)\right].
\end{equation}

In both cases, we consider a grid of values to tune $M$ which has a lower bound of 20\% of the entire training dataset, and an upper bound of 70\% of the entire training dataset, as per our recommendations given in Section 3. We choose 70\% as the upper bound, which we recommended if the interest lies in calibration, rather than the stricter 50\% in this data analysis just as a proof of concept, to check where the optimal $M$ values lie in the grid of values considered under the various values of $\alpha$. 

We begin by displaying the results obtained when using Equation (\ref{eq: newLoss}) as the loss function. A dot plot showing the resulting optimal $M$ proportions is shown in Figure 5(a), and is consistent with what we expected: since both moderate calibration and the lack of spread perform well when $M$ is low, the optimal $M$ value would be low regardless of the choice of $\alpha$. The optimal $M$ proportion under all values of $\alpha$ is 0.29, except $\alpha$ = 0.9 where the optimal $M$ proportion is 0.20. 

\begin{figure*}[h!]
    \centering
        \subfloat[$L^*$]{%
            \includegraphics[width=.5\linewidth]{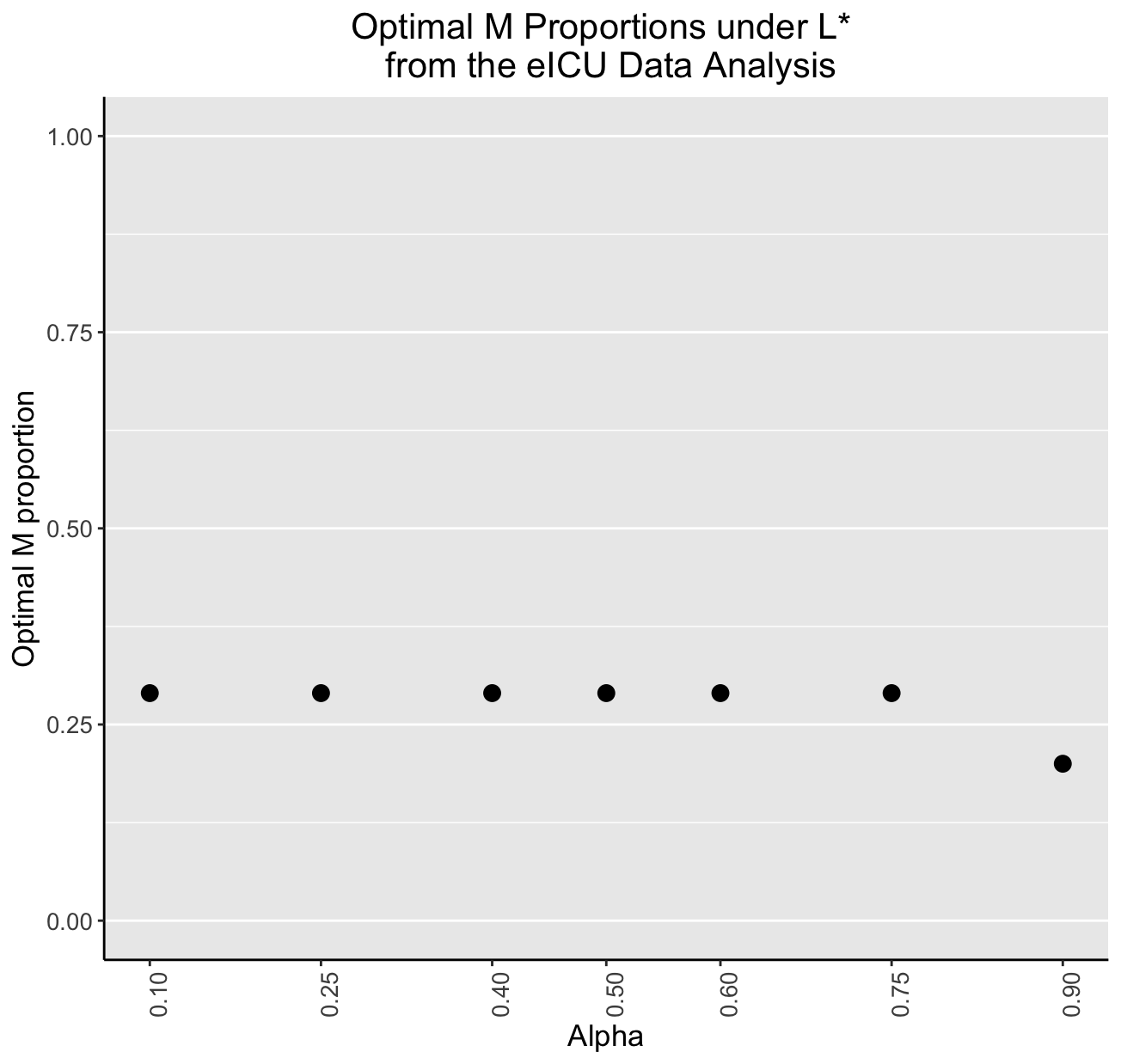}%
        } 
        \subfloat[$L^{**}$]{ 
            \includegraphics[width=.5\linewidth]{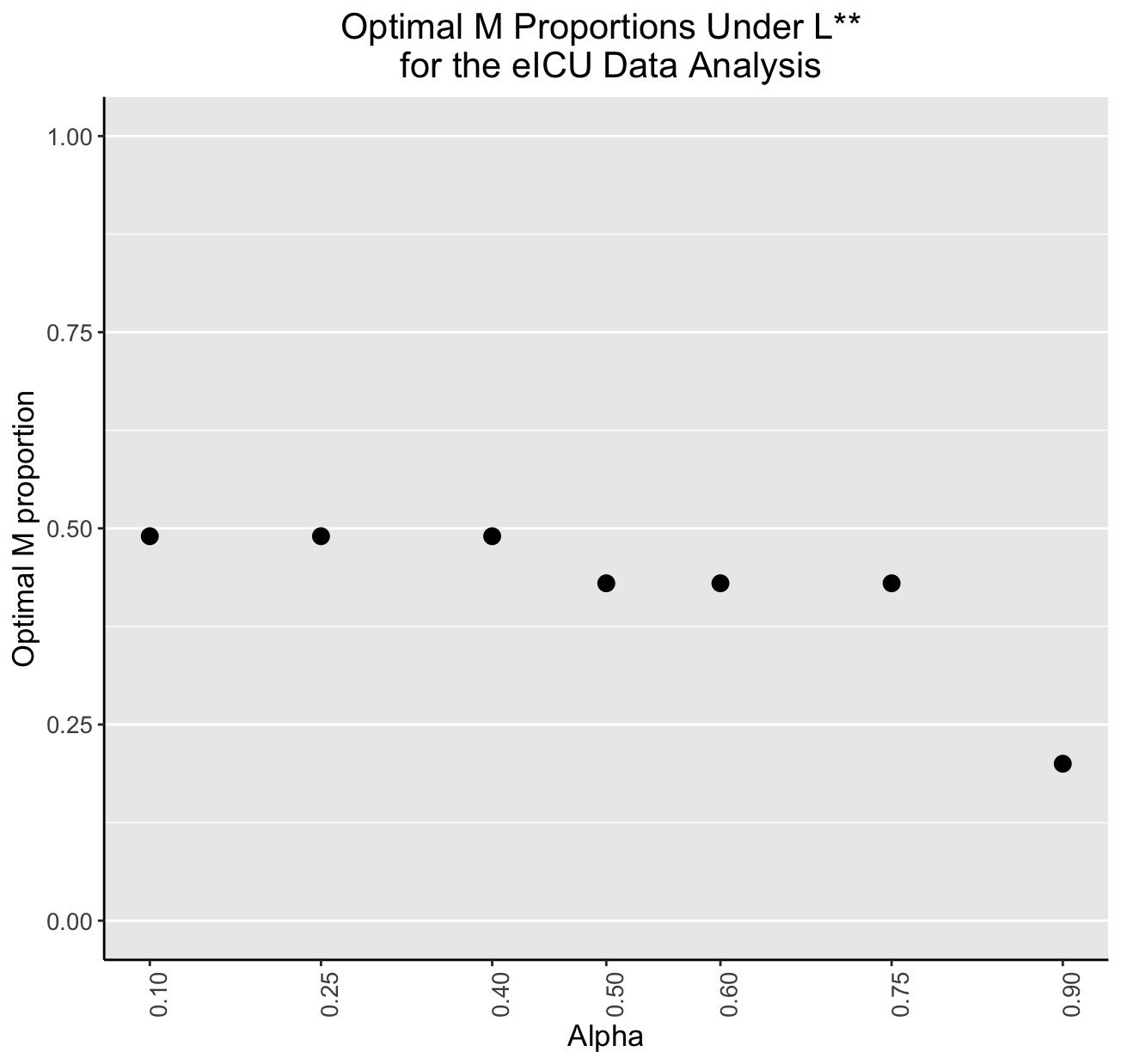}%
        } \\
        \caption{Optimal $M$ proportions found using $L^*$ from Equation (5), and $L^{**}$ from Equation (8), for varying values of $\alpha$ found when tuning $M$ during the training/testing step of applying the proposed algorithm to the eICU dataset.} 
        \label{fig: optimalLin}
    \end{figure*}

Following the training/testing step, we used the optimal proportions shown in Figure 5(a) in the validation step and assessed the performance of the resulting models. The performance of the PPMs under the seven $\alpha$ values, and the performance of the model which uses the entire available training data, is shown in Table 2. Here, we report the BCa confidence interval and the bootstrap standard error in addition to the point estimates of the AUPRC, the lack of spread, CITL, calibration slope, and the ICI. We note that under both the lack of spread measure and the ICI, both of which were included in the loss function used to tune the size of subpopulation, a model fit on a similar subpopulation performs better than one fit on the entire training dataset (ie. Proportion = 1.0). Specifically, when $\alpha = 0.1$, the resulting optimal $M$ proportion is 0.29, which is the best performing PPM in terms of lack of spread, with the measure valued at 0.063, compared to the full model, which has a lack of spread value of 0.071. This is expected as low values of $\alpha$ result in a larger emphasis on the discrimination term in the loss function. When $\alpha = 0.9$, thus giving a larger emphasis on calibration in the loss function, the optimal $M$ proportion is 0.2 and the value of the ICI corresponding to that PPM is 0.013 which is the best moderately calibrated model, which might seem somewhat counter-intuitive with the high $\alpha$, i.e., larger emphasis on calibration, at least based on the simulation results. This said, the difference between the best and worst moderately calibrated model is 0.003, suggesting these small differences are in line with what was found in Figure 4 when investigating the relationship between $M$ and the ICI. 

The AUPRC, which was not considered when tuning $M$ but is a more commonly used measure of discrimination than lack of spread, is the best when the model is fit to the entire training dataset. Under the entire training dataset, the AUPRC is valued at 0.478, while under the optimal $M$ proportion when $\alpha = 0.1$, the resulting PPM has an AUPRC value of 0.475. Based on the investigations shown in Figure 4, the optimal $M$ proportion under the AUPRC is expected to be approximately 0.5, with small differences between the AUPRC of models fit on subpopulations with sizes larger than 50\% of the training data. We explore the changes in optimal $M$ proportion when changing the measure of discrimination in the loss function from the lack of spread to the AUPRC later in this section. 

Under the CITL, the best performing model is the one corresponding to an $M$ proportion of 0.29. Although the point estimates of the CITL for the PPM fit on 29\% and 20\% of the entire training dataset are the same, the BCa confidence interval of the former shows better calibration than the latter. The model with the worst CITL is the full model. Similarly, the model with the best weak calibration, measured by the calibration slope, is the one corresponding to an $M$ proportion of 0.29, while the worst model is the full model. 

In this example, the best performing model would be the one corresponding to any of the $\alpha$ values except $\alpha$ = 0.9, since almost all performance measures are better under an $M$ proportion of 0.29. Further, even though this is not the optimal model under the ICI, the difference between the point estimates of this model and the one which is optimal is 0.002, so we sacrifice little in terms of moderate calibration. For simplicity, given the results we observed, we will say that the best performing model corresponds to $\alpha$ = 0.5, which had a resulting optimal $M$ proportion of 0.29.

\begin{table}[] 
\caption{Cardiac Validation Under $L^*$}
\resizebox{\textwidth}{!}{\begin{tabular}{|l|l|l|l|l|l|l|}
\hline
\multicolumn{1}{|c|}{alpha} & \multicolumn{1}{c|}{Proportion}                                           & \multicolumn{1}{c|}{AUPRC}                                           & \multicolumn{1}{c|}{\begin{tabular}[c]{@{}c@{}}Lack of \\ Spread\end{tabular}} & \multicolumn{1}{c|}{CITL}                                            & \multicolumn{1}{|c|}{Slope} & \multicolumn{1}{c|}{ICI}                                             \\ \hline
0.1                         & 0.29                              & \begin{tabular}[c]{@{}l@{}}0.475 (0.060) \\ (0.363, 0.607)\end{tabular} & \begin{tabular}[c]{@{}l@{}}0.063 (0.004) \\ (0.059, 0.073) \end{tabular}           & \begin{tabular}[c]{@{}l@{}}0.021 (0.006) \\ (0.017, 0.023)\end{tabular} & \begin{tabular}[c]{@{}l@{}}1.015 (0.092) \\ (0.914, 1.129)\end{tabular} & \begin{tabular}[c]{@{}l@{}}0.015 (0.005) \\ (0.006, 0.016)\end{tabular} \\ \hline
0.25                       & 0.29                              & \begin{tabular}[c]{@{}l@{}}0.475 (0.060) \\ (0.363, 0.607)\end{tabular} & \begin{tabular}[c]{@{}l@{}}0.063 (0.004) \\ (0.059, 0.073) \end{tabular}           & \begin{tabular}[c]{@{}l@{}}0.021 (0.006) \\ (0.017, 0.023)\end{tabular} & \begin{tabular}[c]{@{}l@{}}1.015 (0.092) \\ (0.914, 1.129)\end{tabular} & \begin{tabular}[c]{@{}l@{}}0.015 (0.005) \\ (0.006, 0.016)\end{tabular} \\ \hline
0.4                         & 0.29                              & \begin{tabular}[c]{@{}l@{}}0.475 (0.060) \\ (0.363, 0.607)\end{tabular} & \begin{tabular}[c]{@{}l@{}}0.063 (0.004) \\ (0.059, 0.073) \end{tabular}           & \begin{tabular}[c]{@{}l@{}}0.021 (0.006) \\ (0.017, 0.023)\end{tabular} & \begin{tabular}[c]{@{}l@{}}1.015 (0.092) \\ (0.914, 1.129)\end{tabular} & \begin{tabular}[c]{@{}l@{}}0.015 (0.005) \\ (0.006, 0.016)\end{tabular} \\ \hline
0.5                         & 0.29                              & \begin{tabular}[c]{@{}l@{}}0.475 (0.060) \\ (0.363, 0.607)\end{tabular} & \begin{tabular}[c]{@{}l@{}}0.063 (0.004) \\ (0.059, 0.073) \end{tabular}           & \begin{tabular}[c]{@{}l@{}}0.021 (0.006) \\ (0.017, 0.023)\end{tabular} & \begin{tabular}[c]{@{}l@{}}1.015 (0.092) \\ (0.914, 1.129)\end{tabular} & \begin{tabular}[c]{@{}l@{}}0.015 (0.005) \\ (0.006, 0.016)\end{tabular} \\ \hline
0.6                        & 0.29                              & \begin{tabular}[c]{@{}l@{}}0.475 (0.060) \\ (0.363, 0.607)\end{tabular} & \begin{tabular}[c]{@{}l@{}}0.063 (0.004) \\ (0.059, 0.073) \end{tabular}           & \begin{tabular}[c]{@{}l@{}}0.021 (0.006) \\ (0.017, 0.023)\end{tabular} & \begin{tabular}[c]{@{}l@{}}1.015 (0.092) \\ (0.914, 1.129)\end{tabular} & \begin{tabular}[c]{@{}l@{}}0.015 (0.005) \\ (0.006, 0.016)\end{tabular} \\ \hline
0.75                        & 0.29                              & \begin{tabular}[c]{@{}l@{}}0.475 (0.060) \\ (0.363, 0.607)\end{tabular} & \begin{tabular}[c]{@{}l@{}}0.063 (0.004) \\ (0.059, 0.073) \end{tabular}           & \begin{tabular}[c]{@{}l@{}}0.021 (0.006) \\ (0.017, 0.023)\end{tabular} & \begin{tabular}[c]{@{}l@{}}1.015 (0.092) \\ (0.914, 1.129)\end{tabular} & \begin{tabular}[c]{@{}l@{}}0.015 (0.005) \\ (0.006, 0.016)\end{tabular} \\ \hline
0.9                         & 0.20                              & \begin{tabular}[c]{@{}l@{}}0.441 (0.060) \\ (0.358, 0.529)\end{tabular} & \begin{tabular}[c]{@{}l@{}}0.066 (0.004) \\ (0.058, 0.075)\end{tabular}           & \begin{tabular}[c]{@{}l@{}}0.021 (0.007) \\ (0.020, 0.025)\end{tabular} & \begin{tabular}[c]{@{}l@{}}0.981 (0.090) \\ (0.905, 1.057)\end{tabular} & \begin{tabular}[c]{@{}l@{}}0.013 (0.007) \\ (0.007, 0.018)\end{tabular} \\ \hline
                            & 1.0                              & \begin{tabular}[c]{@{}l@{}}0.478 (0.060) \\ (0.375, 0.594)\end{tabular} & \begin{tabular}[c]{@{}l@{}}0.071 (0.004) \\ (0.062, 0.076)\end{tabular}           & \begin{tabular}[c]{@{}l@{}}0.022 (0.006) \\ (0.020, 0.026)\end{tabular} & \begin{tabular}[c]{@{}l@{}}1.024 (0.099) \\ (0.878, 1.198)\end{tabular} & \begin{tabular}[c]{@{}l@{}}0.016 (0.005) \\ (0.005, 0.027)\end{tabular} \\ \hline
\end{tabular}}
\label{table: cardiac2}
\end{table} 

We now discuss the results obtained using the second loss function, $L^{**}$, described in Equation (\ref{eq: auprcLoss}), in applying our algorithm to the eICU cardiac dataset. Recall that this loss function consists of the AUPRC and the ICI as measures of discrimination and calibration, respectively. Based on the relationship between $M$ and the AUPRC seen in Figure 4, we expected the optimal $M$ proportion under $L^{**}$ to be much higher for low values of $\alpha$ compared to what was optimal using $L^*$ as the loss function, as shown in Figure 5(a). The results of this investigation are shown in Figure 5(b), where our hypothesis is shown to be correct. Unlike the results using the lack of spread as the measure of discrimination, here, for low values of $\alpha$, the optimal $M$ proportion is 0.49. The optimal $M$ proportion remains at 0.49 until we set $\alpha$ to be 0.5, which places equal weight on both the AUPRC and the ICI in the loss function, where the proportion then decreases slightly to 0.43. Then, for $\alpha = 0.9$, the optimal $M$ proportion is 0.2, which is exactly what was found under the previous loss function. This is not surprising, as in both loss functions considered, we used the same measure of calibration. These results match the relationships between $M$ and the AUPRC and $M$ and the ICI we observed in Figures 4(a) and (e), respectively. 

Using the optimal $M$ proportion found under each value of $\alpha$, we then validated the results and display our findings in Table 3. We immediately notice that the AUPRC of the model corresponding to $\alpha = 0.1$ has improved compared to the results using the prior loss function, shown in Table 2, and it outperforms the AUPRC of the full model. The resulting AUPRC performance confirms the relationship between $M$ and the AUPRC we observed in Figure 4, where the model with the optimal AUPRC is fit on approximately 50\% of the full training dataset, but the difference in AUPRC values between the optimal and full models is very small. Specifically, here, the optimal model, which is fit on a similar subpopulation made up of 49\% of the patients from the full training dataset, has an AUPRC value of 0.481, while the full model has an AUPRC value of 0.478. As expected, the lack of spread of the model is not optimal under an $M$ proportion of 0.49, with a value that is 0.006 units higher than what was displayed in Table 2 when we set $\alpha = 0.1$, however, the performance is still better than the full model, which, again, confirms the relationships seen in Figure 4. 

As we did not change the measure of calibration used to tune the size of subpopulation in the loss function, we do not see many changes in the calibration of these models. The model with the best moderate calibration is still the model corresponding to an $M$ proportion of 0.2, which was determined when we put most of the weight in the loss function on the calibration term (ie. $\alpha = 0.9$). The model with the best mean calibration, measured through the CITL, is the one which is fit on 43\% on the full training dataset, while the model with the best weak calibration is the one which is fit on 20\% of the training dataset.  

The overall optimal model using $L^{**}$ as the loss function to tune $M$ is not as clear of a decision as we observed when using $L^*$ as the loss function. Here, the user would have to make a decision as to what they value more and what measure they are willing to sacrifice. As the differences between measures are very small, these sacrifices will in turn be small in size. We define the optimal model of this study as the one which results in the smallest sacrifices in terms of performance, which, in this case, corresponds to $\alpha = 0.1$, $\alpha = 0.25$ or $\alpha = 0.4$, where all three result in an optimal $M$ proportion of 0.49. Without loss of generality, we conclude that the model corresponding to $\alpha = 0.4$ is optimal, where this model has the best discrimination in terms of the AUPRC, and the lack of spread is 0.001 units less than its optimal value, the CITL is 0.001 units smaller than its optimal value, the calibration slope is 0.005 absolute units smaller than its optimal value and the ICI is 0.003 units smaller than its optimal value across all eight models considered. 

\begin{table}[] 
\caption{Cardiac Validation Under $L^{**}$}
\resizebox{\textwidth}{!}{\begin{tabular}{|l|l|l|l|l|l|l|}
\hline
\multicolumn{1}{|c|}{alpha} & \multicolumn{1}{c|}{Proportion}                                           & \multicolumn{1}{c|}{AUPRC}                                           & \multicolumn{1}{c|}{\begin{tabular}[c]{@{}c@{}}Lack of \\ Spread\end{tabular}} & \multicolumn{1}{c|}{CITL}                                            & \multicolumn{1}{|c|}{Slope} & \multicolumn{1}{c|}{ICI}                                             \\ \hline
0.1                         & 0.49                              & \begin{tabular}[c]{@{}l@{}}0.481 (0.060) \\ (0.375, 0.595)\end{tabular} & \begin{tabular}[c]{@{}l@{}}0.069 (0.004) \\ (0.059, 0.075) \end{tabular}           & \begin{tabular}[c]{@{}l@{}}0.022 (0.006) \\ (0.020, 0.026)\end{tabular} & \begin{tabular}[c]{@{}l@{}}1.024 (0.099) \\ (0.878, 1.198)\end{tabular} & \begin{tabular}[c]{@{}l@{}}0.016 (0.006) \\ (0.005, 0.027)\end{tabular} \\ \hline
0.25                       & 0.49                              & \begin{tabular}[c]{@{}l@{}}0.481 (0.060) \\ (0.375, 0.595)\end{tabular} & \begin{tabular}[c]{@{}l@{}}0.069 (0.004) \\ (0.059, 0.075) \end{tabular}           & \begin{tabular}[c]{@{}l@{}}0.022 (0.006) \\ (0.020, 0.026)\end{tabular} & \begin{tabular}[c]{@{}l@{}}1.024 (0.099) \\ (0.878, 1.198)\end{tabular} & \begin{tabular}[c]{@{}l@{}}0.016 (0.006) \\ (0.005, 0.027)\end{tabular} \\ \hline
0.4                         & 0.49                              & \begin{tabular}[c]{@{}l@{}}0.481 (0.060) \\ (0.375, 0.595)\end{tabular} & \begin{tabular}[c]{@{}l@{}}0.069 (0.004) \\ (0.059, 0.075) \end{tabular}           & \begin{tabular}[c]{@{}l@{}}0.022 (0.006) \\ (0.020, 0.026)\end{tabular} & \begin{tabular}[c]{@{}l@{}}1.024 (0.099) \\ (0.878, 1.198)\end{tabular} & \begin{tabular}[c]{@{}l@{}}0.016 (0.006) \\ (0.005, 0.027)\end{tabular} \\ \hline
0.5                         & 0.43                              & \begin{tabular}[c]{@{}l@{}}0.475 (0.060) \\ (0.369, 0.593)\end{tabular} & \begin{tabular}[c]{@{}l@{}}0.068 (0.004) \\ (0.060, 0.074) \end{tabular}           & \begin{tabular}[c]{@{}l@{}}0.021 (0.006) \\ (0.020, 0.030)\end{tabular} & \begin{tabular}[c]{@{}l@{}}0.848 (0.097) \\ (0.660, 0.979)\end{tabular} & \begin{tabular}[c]{@{}l@{}}0.014 (0.005) \\ (0.005, 0.023)\end{tabular} \\ \hline
0.6                         & 0.43                              & \begin{tabular}[c]{@{}l@{}}0.475 (0.060) \\ (0.369, 0.593)\end{tabular} & \begin{tabular}[c]{@{}l@{}}0.068 (0.004) \\ (0.060, 0.074) \end{tabular}           & \begin{tabular}[c]{@{}l@{}}0.021 (0.006) \\ (0.020, 0.030)\end{tabular} & \begin{tabular}[c]{@{}l@{}}0.848 (0.097) \\ (0.660, 0.979)\end{tabular} & \begin{tabular}[c]{@{}l@{}}0.014 (0.005) \\ (0.005, 0.023)\end{tabular} \\ \hline
0.75                        & 0.43                              & \begin{tabular}[c]{@{}l@{}}0.475 (0.060) \\ (0.369, 0.593)\end{tabular} & \begin{tabular}[c]{@{}l@{}}0.068 (0.004) \\ (0.060, 0.074) \end{tabular}           & \begin{tabular}[c]{@{}l@{}}0.021 (0.006) \\ (0.020, 0.030)\end{tabular} & \begin{tabular}[c]{@{}l@{}}0.848 (0.097) \\ (0.660, 0.979)\end{tabular} & \begin{tabular}[c]{@{}l@{}}0.014 (0.005) \\ (0.005, 0.023)\end{tabular} \\ \hline
0.9                         & 0.20                              & \begin{tabular}[c]{@{}l@{}}0.441 (0.060) \\ (0.358, 0.529)\end{tabular} & \begin{tabular}[c]{@{}l@{}}0.066 (0.004) \\ (0.058, 0.075)\end{tabular}           & \begin{tabular}[c]{@{}l@{}}0.021 (0.007) \\ (0.020, 0.025)\end{tabular} & \begin{tabular}[c]{@{}l@{}}0.981 (0.009) \\ (0.905, 1.057)\end{tabular} & \begin{tabular}[c]{@{}l@{}}0.013 (0.007) \\ (0.007, 0.018)\end{tabular} \\ \hline
                            & 1.0                              & \begin{tabular}[c]{@{}l@{}}0.478 (0.060) \\ (0.375, 0.594)\end{tabular} & \begin{tabular}[c]{@{}l@{}}0.071 (0.004) \\ (0.062, 0.076)\end{tabular}           & \begin{tabular}[c]{@{}l@{}}0.022 (0.006) \\ (0.020, 0.026)\end{tabular} & \begin{tabular}[c]{@{}l@{}}1.024 (0.099) \\ (0.878, 1.198)\end{tabular} & \begin{tabular}[c]{@{}l@{}}0.016 (0.005) \\ (0.005, 0.027)\end{tabular} \\ \hline
\end{tabular}}
\label{table: cardiacloss2}
\end{table} 

\section{Discussion}
\label{s: discussion}

We introduced a general loss function on which to tune $M$ when fitting a personalized predictive model (PPM) which allows not only the flexibility of weighting one predictive model performance measure over another, but also allows flexibility in choosing which measures of model discrimination and calibration to be considered when optimizing the size of subpopulation. Further work is necessary to verify if this loss function is a proper scoring rule. Further, we investigated the relationship between $M$ and various measures of model discrimination and calibration, and gave recommendations on bounds for the grid of values on which to tune $M$ with the goal of reducing computation time.

We simulated 12 datasets that differed by their true outcome model, and analyzed the relationships between $M$ and six performance measures: AUROC, AUPRC, lack of spread, CITL, calibration slope and the ICI. In all cases, $M$ had a negatively proportional relationship with measures of discrimination, aside from the lowest values of $M$ when using the AUROC and the AUPRC as performance measures, so a relatively low value of $M$ will result in a PPM with relatively better discriminative ability than a model fit on a larger subpopulation. In many cases, this low value is good enough, if not the best, for model calibration as well. The optimal $M$ was often very similar under the calibration slope and the ICI, which are measures of weak calibration and moderate calibration, respectively. This agreement between the optimal $M$ values did not always extend to the CITL, a measure of mean calibration, the lowest calibration measure in the calibration hierarchy \cite{van2016calibration} which had an optimal $M$ that was sometimes different than that under weak and moderate calibration. Based on the results from this simulation study, we recommend that the grid of values on which to tune $M$ should be bounded below by 20\% of the training data, and bounded above by either 50\% or 70\% of the training data, depending on whether there is greater interest on model discrimination or calibration. A lower bound of 20\% ensures that the sample size in the validation step will not be too small, and corresponds to what we see as optimal under discrimination in most cases we considered. An upper bound of 50\% when there is equal interest in discrimination and calibration, or greater interest in discrimination (ie. $\alpha \leq 0.5$) tightens the bound and decreases computation time while still likely capturing the optimal $M$ value under discrimination. An upper bound of 70\% when greater interest lies in calibration (ie. $\alpha > 0.5$) increases the chance of capturing the optimal $M$ under calibration if that value is large, while ensuring that we do not have too much deterioration in discriminative ability. Since a low value of $M$ is good, if not the best, in many cases in terms of calibration, we can justify not having the entire size of the training dataset as the upper bound on the grid of values on which to tune $M$. By bounding the grid of values, computation time can be cut down significantly, which is important for when these methods are taken into practice, as PPM algorithms are computationally intensive, given that a unique model is fit for each individual in the test data. The cases we considered here are not part of an exhaustive list, and these recommendations on bounds are based on the 12 cases we did consider. Other datasets may differ in characteristics that may lead to the consideration of a wider grid, however, our recommendations serve, at the very least, as a useful starting point. Future work includes investigating these relationships under more datasets, as well as more models (ie. random forest, or more complex machine learning models), especially the relationships between $M$ and calibration, in order to propose bounds that are more robust.

Through an extensive simulation study, we tested our proposed general loss function which consisted of the lack of spread as a measure of discrimination, and the ICI as a measure of calibration. We tested multiple values of $\alpha$ to find the optimal size of subpopulation when fitting a PPM and compared these results to the relationships investigated in Section 3.2. As we increased the weight put on the ICI term in the loss function, we found that the resulting optimal $M$ matched what was optimal under the ICI measure in the relationships displayed in Figures 1 and 2. In future work, we wish to explore whether there are certain combinations of calibration and discrimination measures that can be used in the loss function shown in Equation (\ref{eq: generalLoss}) that lead to improved loss and ultimately better predictive performance. 

We also applied the proposed algorithm to a dataset from the eICU database to predict the mortality of patients with diseases of the circulatory system. We considered two different loss functions which both had the ICI as the measure of calibration, but differed in the defined measure of discrimination, which was either the lack of spread, as was done in the simulation study, or the AUPRC, a more common measure of discrimination, particularly when the prevalence of the outcome is small, which was the case for our data analysis. In the first loss function considered, where the discrimination in the loss function was the lack of spread, the optimal $M$ proportion under all values of $\alpha$ was low, specifically, it was 0.29 for all values except for $\alpha = 0.9$, where the optimal $M$ was 0.2. In the second loss function considered, there was more variability in the optimal $M$ proportions across $\alpha$ values, and the optimal $M$ proportions when discrimination was emphasized (ie. when $\alpha$ was a value close to 0), was a much higher value at 0.49. These results coincided with the relationships between $M$ and spread, as well as $M$ and the AUPRC which we saw in Figure 4: very low values of $M$ correspond to low values of lack of spread, while moderate values of $M$ correspond to higher values of the AUPRC. We fit PPMs to the hold-out dataset using the optimal $M$ proportions found in the training/testing step and calculated BCa bootstrap confidence intervals for each of the point estimates of the performance measures in both cases. Under the first loss function, we concluded that the optimal model is the one corresponding to $\alpha = 0.5$, which resulting in an optimal $M$ proportion of 0.29, where all performance measures were optimized except the ICI, which performed optimally under $\alpha = 0.2$ (ie. $M_{proportion} = 0.2$). The model fit to the full training dataset had the worst performance across all performance measures, except under the AUPRC, where it performed slightly better. When using the AUPRC as the discrimination in the loss function, we concluded that the optimal model corresponded to $\alpha = 0.4$ (ie. $M_{proportion} = 0.49$), as choosing this model resulted in the smallest sacrifices made in calibration performance. Further, the full model, under the second loss function, was the worst performing model across all performance measures considered. Through this data analysis, we showed how the optimal model differs when changing the choice of performance measures to use in the loss function.  

This study has a few limitations. Computational intensity continues to be the greatest challenge to overcome before taking these algorithms into practice. We introduced some coding techniques to help improve computation time, such as the use of Rcpp \cite{Rcpp} and parallelization. We have also improved on the computation time by introducing bounds to the grid of values used to tune $M$. However, this is still not sufficient, especially if these methods are to be used in real time, for example in an intensive care unit, where decisions must be made very quickly. As mentioned prior, we focus our efforts on a relatively basic model, logistic regression, to explore these relationships between the size of subpopulation, and the effects of patient similarity methods on prediction, under a simple case. Next steps of this work include investigations under more sophisticated machine learning models, that is, to see what the optimal $M$ values would be using the same data but under different models. We also would like to investigate the methods proposed on a broader family of non-linear models. However, this exploration needs to await further developments on computational efficiencies for the proposed algorithm.

\clearpage

\bibliographystyle{apalike}

\end{document}